\documentclass{article}
\pdfoutput=1  

\usepackage{graphicx}
\usepackage{amssymb}
\usepackage{cite}

\newcommand{\eq}[1]{Eq.~(\ref{eq.#1})} 
\newcommand{\fig}[1]{Fig.~\ref{fig.#1}}

\newcommand{\tbl}[1]{Table~\ref{table.#1}}

\newcommand{\sect}[1]{Section~\ref{sect.#1}}
\newcommand{\sectA}[1]{Appendix~\ref{sect.#1}}
\newcommand{\sectlabel}[1]{\label{sect.#1}}
\newcommand{\eqlabel}[1]{\label{eq.#1}}
\newcommand{\figlabel}[1]{\label{fig.#1}}
\newcommand{\tbllabel}[1]{\label{table.#1}}

\newcommand{\code}[1]{\mbox{\texttt{#1}}} 

\newcommand{\BoltzmannConstant}{\ensuremath{k_{\mbox{\scriptsize B}}}}

\newcommand{\Lthermal}{\ensuremath{L_{\mbox{\scriptsize th}}}}
\newcommand{\Wthermal}{\ensuremath{\omega_{\mbox{\scriptsize th}}}}

\newcommand{\Kdrag}{\ensuremath{k_{\mbox{\scriptsize rd}}}}

\newcommand{\Edissipated}{\ensuremath{E_{\mbox{\scriptsize dissipated}}}}
\newcommand{\Pdissipated}{\ensuremath{P_{\mbox{\scriptsize dissipated}}}}

\newcommand{\meter}{\mbox{m}}
\newcommand{\nanometer}{\mbox{nm}}
\newcommand{\second}{\mbox{s}}
\newcommand{\nanosecond}{\mbox{ns}}
\newcommand{\picosecond}{\mbox{ps}}
\newcommand{\femtosecond}{\mbox{fs}}
\newcommand{\kilogram}{\mbox{kg}}

\newcommand{\joule}{\mbox{J}}
\newcommand{\kJmol}{\mbox{kJ/mol}}
\newcommand{\calmol}{\mbox{cal/mol}}

\newcommand{\gigahertz}{\mbox{GHz}}
\newcommand{\kgmsq}{\mbox{kg m}^2}
\newcommand{\kgmsqpers}{\mbox{kg\,m${}^2$/s}}
\newcommand{\fluctuation}{\mbox{kg\,m${}^2$/s/$\sqrt{\nanosecond}$}}
\newcommand{\Kelvin}{\mbox{K}}
\newcommand{\piconewton}{\mbox{pN}}
\newcommand{\radian}{\mbox{rad}}

\newcommand{\widefigwidth}{5.0in}
\newcommand{\figwidth}{3.5in}

\begin{document}

\title{Evaluating the Friction of Rotary Joints in Molecular Machines}
\author{Tad Hogg\\
{\small Institute for Molecular Manufacturing}\\{\small Palo Alto, CA }
\and Matthew S. Moses \\
{\small Independent Consultant}\\{\small Lafayette, CO}
\and Damian G. Allis\\
{\small Department of Chemistry}\\{\small Syracuse University}\\{\small Syracuse, NY}
}

\maketitle
\begin{abstract}
A computationally-efficient method for evaluating friction in molecular rotary bearings is presented. This method estimates drag from fluctuations in molecular dynamics simulations via the fluctuation-dissipation theorem. This is effective even for simulation times short compared to a bearing's energy damping time and for rotation speeds comparable to or below typical thermal values. We apply this method to two molecular rotary bearings of similar size at $300\,\Kelvin$: previously studied nested (9,9)/(14,14) double-walled carbon nanotubes and a hypothetical rotary joint consisting of single acetylenic bonds in a rigid diamondoid housing. 
The acetylenic joint has a rotational frictional drag coefficient of $2 \times 10^{-35}\,\kgmsqpers$. 
The friction for the nested nanotubes is 120 times larger, comparable to values reported by previous studies.
This fluctuation-based method could evaluate dissipation in a variety of molecular systems with similarly rigid and symmetric bearings.

\end{abstract}

\section{Introduction}\sectlabel{introduction}

Bonds between carbon atoms are among the strongest chemical bonds that are free to rotate, subject only to the steric and electrostatic interactions between bound substituents. Hydrocarbons and organic molecules often have small potential energy barriers to this rotation, allowing a molecule to freely change configuration as constituent atoms rotate~\cite{pitzer51,kottas05}. 
The primary advantages of these molecular rotary joints, apart from their incredibly small size, are very low friction and zero wear. 
This phenomenon has enabled low-dissipation rotary motion in a number of recent nanoscale structures. 
For example, molecular machines with rotating or sliding joints have been created by various techniques, including traditional chemical synthesis~\cite{kottas05, shirai05, ogi10, gimzewski98, pekker05, perera13, zhang16, manzano09,balzani2000artificial,browne2006making,champin2007transition}, manipulation of carbon nanotubes~\cite{cumings00, zhang13}, manipulation of graphene~\cite{koren15}, and re-purposing biological motors~\cite{nicolau16}. 
In theory, molecular machines using carbon bonds include planetary gearboxes~\cite{drexler92,cagin98}, flywheels for energy storage~\cite{freitas99}, and mechanical computers~\cite{merkle16}. Such nanomachines will require large-scale atomically precise manufacturing, which is a challenging problem that is as yet unsolved in the general case.

A key performance measure for molecular machines is their energy dissipation. Unfortunately, experimental measurements are often difficult to obtain, and impossible for molecular machines that cannot yet be fabricated. Computational studies are a reasonable alternative for evaluating and optimizing designs for various applications.
Conceptually, estimating dissipation for a molecular machine is straightforward: simulate the machine's behavior over enough time for energy introduced into the machine's operational degrees of freedom to irreversibly transfer to many other degrees of freedom.
Unfortunately, such computational studies are challenging for machines with many atoms and when the machine's operation is only weakly coupled to other atoms, leading to low friction and long damping times. Simulating such machines long enough to observe this damping may not be feasible, even for machines operating in vacuum, where the absence of gas or solvent molecules greatly reduces the computational cost.
Moreover, even if simulating a single design is feasible, identifying low-power designs may require examining many machine variants, leading to a computationally costly overall design task.

One approach to reduce the computation time is to simulate the machine at speeds well above the typical thermal speed of the machine's motion. In this case, thermal fluctuations are relatively small, allowing definite estimates of damping directly from the change in speed, even over times short compared to the damping time. For example, this approach has been applied to determine friction in nested nanotubes~\cite{cook13}.
However, this method may not apply to large molecular machines intended to operate near or below thermal speeds. This is because the machine may behave very differently at low and high speeds. For instance, machines at high speeds could involve forces strong enough to significantly distort components from their low-speed geometries or even break chemical bonds. More generally, high-speed operation may excite high-frequency vibrational modes that would not be relevant at intended operational speeds, thereby significantly overestimating the damping the machine would experience at its intended operating speed.

To address this challenge, this paper examines the effectiveness of an alternate method~\cite{torres06}: using fluctuations in the simulated behavior to infer friction via the fluctuation-dissipation theorem. In contrast to measuring the damping directly from the decreasing energy during the simulation, fluctuations are evident even in simulations over times much shorter than the machine's damping time and when operating at speeds comparable to or less than thermal speeds. 

This paper evaluates this fluctuation-based method for two examples. The first, rotating nested nanotubes, has been studied extensively and allows comparison with direct estimation of dissipation from the decrease of the energy. In this case, the relatively small number of atoms and high dissipation make both the direct and fluctuation-based methods feasible with modest computational cost. 

The second example is a hypothetical molecular rotor using acetylenic bonds. These bonds give exceptionally low rotary friction, and thereby would require simulating the machines over long times to reliably measure dissipation directly. Moreover, including enough atoms around those bonds to capture interactions with the rotor's housing results in a significantly larger number of atoms than used for the nanotube example. Thus this rotor is an example where the fluctuation-based method could provide a significant benefit.

Specifically, after discussing related studies, \sect{drag estimate} describes the fluctuation-based method. We then present the two examples. The concluding section discusses future directions for applying the method and its possible limitations. The appendix provides details of the simulations.

\section{Related Work}\sectlabel{related work}

Several studies have used molecular dynamics (MD) simulations to investigate friction in nanomachines with sliding or rotating components. 

In \cite{cook13}, four different MD simulation configurations using the LAMMPS code and AIREBO force field were used to obtain consistent friction estimates for a rotating nested nanotube bearing. Other dissipation estimates for rotating nested nanotube bearings have been determined using intralayer interactions based on the Brenner potential \cite{brenner90} with interlayer interactions based on the Kolmogorov-Crespi registry-dependent potential~\cite{zhang04}, and with the COMPASS force field~\cite{kan14}. Dissipation estimates for \emph{linear} sliding nested nanotube bearings have been determined using a custom force field and custom numerical simulator in \cite{servantie06, servantie06b}, and a custom force field based on \cite{quo91} and the r-RESPA integrator in \cite{rivera05}.

A MD model of experimentally realized graphene-on-graphene sliding is presented in \cite{koren15}. The self-retracting motion of sheared graphene sheets is studied with an in-house MD integrator in \cite{popov11}. Energy dissipation during high speed sliding of graphene sheets is investigated with GROMACS~\cite{hess08} and the DREIDING force field~\cite{mayo90} in~\cite{liu14}.

In addition to the relatively simple cases of nested nanotubes and graphene-graphene sliding, MD studies have evaluated more complex nanomachines. One example is a study of meshing gears made from functionalized carbon nanotubes~\cite{han97} using the Brenner potential~\cite{brenner90}.
In \cite{cagin98} a complex molecular planetary gear mechanism is analyzed using the UFF forecefield and Gasteiger partial charges. In \cite{akimov08} a custom integrator is used with a custom force field based on CHARMM and AMBER force fields, with partial charges determined by \emph{ab initio} calculations, to model the behavior of the experimentally realized ``nanocars"~\cite{shirai05} rolling on a gold surface.

An experimental and MD study measured the energy required to force rotation of a sterically-congested molecular bond in a surface-bound molecule~\cite{loppacher03}. The work to rotate one bond was $5\times 10^{-20}\,\joule$. The energy barrier involved in rotating the bond in this molecule is much larger than that of the rotary joint presented in this paper. All of this energy is not necessarily dissipated since much could be recovered by a properly designed mechanism. Nevertheless, this study provides an upper bound on the energy dissipated, and also indicates the work that would need to be supplied and, ideally, recovered for a single rotation. 

\section{Estimating Drag from Fluctuations} \sectlabel{drag estimate}

A physical system used as a machine distinguishes a few operational dynamic properties from other degrees of freedom of the machine and its environment. 
Conservative forces, derived from potential energy gradients, 
may impede the machine's motion. But they do not by themselves account for dissipation: energies associated with conservative forces can, in principle, be recovered during cyclic operation.
On the other hand, friction is a dissipative force arising from random thermal interactions among the atoms of the machine and its environment. These interactions dissipate organized energy in the machine's motion into heat, and appear as a damping force. The distinction between conservative and dissipative forces is important for evaluating dissipation~\cite{cook13}.

\subsection{Stochastic Model of Rotational Damping}\sectlabel{stochastic model}

This paper considers rotary bearings which, ideally, have a single operational degree of freedom: rotation about a fixed axis. A molecular rotary joint can approximate this behavior~\cite{kottas05}, with additional degrees of freedom providing rapidly fluctuating torques on the rotation, i.e., a thermal bath. These torques produce a drag on the rotor.

We model the effect of these torques with a phenomenological Langevin equation~\cite{gillespie92,lemons02} for the rotor's angular momentum $L$~\cite{kottas05}:
\begin{equation}\eqlabel{L stochastic}
d L = -\gamma L \,dt - \frac{dV}{d\theta}\, dt + \sigma \,dW
\end{equation}
The first term on the right-hand side is a frictional damping torque, with characteristic damping time $1/\gamma$. The second term describes conservative torques, arising from a potential energy $V(\theta)$ depending on the rotor's orientation angle $\theta$. For the machines considered in this paper, torques arising from the potential are relatively small. In the final term, $W$ is a standard Wiener stochastic process~\cite{gillespie92} modeling the fluctuating torques. Operationally,  $\sigma \, dW$ denotes independent normally-distributed random values with zero mean and standard deviation $\sigma \sqrt{\delta t}$ over a time interval $\delta t$.
The rotor's angular position changes as $d\theta/dt =\omega$, where $\omega= L/J$ is the angular velocity and $J$ is the rotor's moment of inertia.
\eq{L stochastic} describes a stochastic process~\cite{gillespie92} and numerical integration~\cite{gillespie92,higham01} allows sampling solutions to this equation. This stochastic process is the rotational analog of a similar model applied to friction associated with linear motion~\cite{persson96,torres06}.

The damping term in \eq{L stochastic} corresponds to a drag torque $\tau=\gamma L = \gamma J \omega$. 
This drag is commonly expressed in terms of the rotational frictional drag coefficient, $\Kdrag$, defined by the relation $\tau \equiv \Kdrag \omega$, so that $\Kdrag = \gamma J$.
This drag dissipates energy at the rate $\Pdissipated = \Kdrag \omega^2$.
The energy dissipated due to this friction when rotating by angle $\phi$ at a uniform speed is $\tau \phi$. If this rotation takes place over time $t$, then $\omega = \phi/t$ and the dissipation is
\begin{equation}\eqlabel{dissipation}
\Edissipated = \Pdissipated t = \Kdrag \omega \phi =  \Kdrag \phi^2/t
\end{equation}


When operating at temperature $T$, the fluctuation-dissipation theorem~\cite{gillespie92,lemons02,kottas05} relates the damping and fluctuation parameters, $\gamma$ and $\sigma$.
The physical mechanism underlying this relation is that damping arises from continual interactions with the environment. Random variations in these interactions provide fluctuating torques that, on average, damp the rotor's motion. However, instead of coming to rest, these torques give the rotor a fluctuating angular momentum that is zero on average but has a nonzero variance denoted by $\Lthermal^2$, where $\Lthermal$ is called the thermal angular momentum.
%
The stochastic process of \eq{L stochastic} has equilibrium variance $\sigma^2/(2\gamma)$~\cite{gillespie92}. 
Thus $\Lthermal^2 = \sigma^2/(2\gamma)$.
The thermal angular momentum is related to temperature through the equipartition theorem~\cite{huang63}, which states the average rotational kinetic energy equals $\BoltzmannConstant T/2$, where $\BoltzmannConstant$ is Boltzmann's constant. Since the rotational kinetic energy of a rotor with angular momentum $L$ is $L^2/(2J)$, the average kinetic energy in equilibrium is $\Lthermal^2/(2 J)$.
Thus $\Lthermal^2/(2 J) = \BoltzmannConstant T/2$, so $\Lthermal = \sqrt{\BoltzmannConstant T J}$. Equating the expressions for $\Lthermal$ from the variance of \eq{L stochastic} and equipartition gives
\begin{equation}\eqlabel{fluctuation dissipation}
\sigma^2 = 2 \BoltzmannConstant T J \gamma
\end{equation}
which relates the size of the fluctuations to the damping constant.

\subsection{Range of Model Validity}\sectlabel{validity}

\eq{L stochastic} is an approximate model of a molecular rotor. Using this model to estimate drag requires determining when the model is an adequate approximation. In particular, the model assumes drag is linearly proportional to angular momentum and the motions of the other degrees of freedom of the machine provide uncorrelated random torques on the rotor's angular momentum.

In general, the drag can have a nonlinear dependence on the speed. Such nonlinearity can be significant at high speeds, e.g., close to the speed of sound in the structure or involving forces large enough to break bonds. On the other hand, at sufficiently low speeds, the linear term dominates in a Taylor expansion of the drag as a function of speed. 
As one comparison, the speed of sound in diamond is about $10^4\,\meter/\second$, whereas the examples in this paper involve rotation of carbon molecules with radius $r\approx 1\,\nanometer$ at up to angular velocities of $\omega=100\,\radian/\nanosecond$, with corresponding speed $r\omega \approx 10^2\,\meter/\second$.

The second assumption, uncorrelated random torques from other degrees of freedom, fails at sufficiently short time scales. For example, brief tilts of the rotor axis or oscillating normal modes of the housing could introduce short correlations in torques applied to the rotor. \eq{L stochastic} assumes that these correlation times are short compared to times relevant for machine operation, which include the damping time. This requires a significant separation in time scales between the rotary degree of freedom and other motions. Molecular machines can achieve this separation with strongly bonded atoms throughout the structure except the parts intended to rotate, which instead have relatively weak interactions.  

Thus we expect the stochastic model to apply to rotors operating at sufficiently low speeds and for sufficiently long times. We identify an appropriate time scale by comparing the standard deviation of changes in the rotor's angular momentum over various time scales $\delta t$. Specifically, consider $\Delta L(t) = L(t+\delta t)-L(t)$, the change in angular momentum starting at a time $t$. From a sample of the angular momenta, we compute $s$, the standard deviation of the changes $\Delta L$ at various times $t$ in this sample, as a function of $\delta t$. Provided $\delta t$ is short compared to the damping time $1/\gamma$ and the potential has only a small effect, the fluctuation term of \eq{L stochastic} dominates the changes, so $s \approx \sigma \sqrt{\delta t}$. In this case, the ratio $s/\sqrt{\delta t}$ from the simulations should be nearly independent of $\delta t$. We evaluate this behavior with a ``fluctuation plot'' of this ratio vs.~$\delta t$. The range of times $\delta t$ achieving this independence indicates the time scale at which the molecular machine behaves approximately as a thermally-driven rotor.

Another check on the model validity is comparing the power spectrum of the angular momentum with that of the stochastic process of \eq{L stochastic} using parameters estimated from simulations of the rotor behavior. Frequencies with a close match indicate corresponding times over which the model is a good approximation. In our cases, we find the time scales determined from power spectra are similar to those from fluctuation plots.

\subsection{Molecular Dynamics Simulations}\sectlabel{MD simulation}

We evaluated rotational drag with molecular dynamics simulations. These simulations have two stages: warmup and spin-down. 

The warmup stage starts by adjusting the initial geometry to an energy minimum without constraints on any part of the structure. Next, we impose boundary conditions, i.e., fix the positions of a specific subset of the atoms.
This simulation starts the system at zero temperature and couples it to a temperature bath at $300\,\Kelvin$. We run the warmup simulation long enough bring the atoms to this temperature.

The spin-down stage starts with a warmed up configuration and adds velocities to the atoms of the rotary component to give it a specified initial angular velocity $\omega_0$. We then simulate the system in isolation, i.e., without coupling to a thermal bath. This is appropriate for the machines considered here, operating in vacuum, because the system is not in intimate contact with a thermal reservoir.

We repeat this procedure, with different random seeds, to obtain multiple samples of the behavior.
\sectA{gromacs details} describes the simulation procedure in more detail.

\subsection{Estimating Drag from Fluctuations in Angular Momentum}

The simplest approach to estimating the damping $\gamma$ is to average angular momenta over multiple simulations, since \eq{L stochastic} implies the average decays exponentially: $\langle L \rangle \sim \exp(-\gamma t)$. However, as discussed in \sect{introduction}, this direct approach may not be feasible for molecular machines containing many atoms and with low dissipation.

On the other hand, fluctuations in the angular momentum are readily apparent in simulations over relatively short times. Thus we can use short simulations to estimate the fluctuation parameter $\sigma$ and then determine the damping $\gamma$ via \eq{fluctuation dissipation}. 
We do so by finding the value of $\sigma$ that maximizes the likelihood of the observed changes in angular momentum over a fixed increment time $\delta t$ according to \eq{L stochastic} and subject to the constraint of \eq{fluctuation dissipation}. The second derivative of the log-likelihood at the maximum gives approximate confidence intervals for the estimated parameters.

In this fluctuation-based method, the choice of $\delta t$ is somewhat arbitrary within a broad range. On the one hand, $\delta t$ must be large enough that \eq{L stochastic} is a good approximation for the rotor, as discussed in \sect{validity}. On the other hand, $\delta t$ must be short compared to the overall simulation time to have enough samples for estimation. As illustrated with the examples discussed below, this approach can estimate damping with simulation times short compared to the rotor's damping time even when operating at speeds comparable to thermal motion, so fluctuations are relatively large. For such situations, the fluctuation-based approach is computationally less costly than running simulations long enough to directly estimating damping with similar confidence.

In general, \eq{L stochastic} does not have a simple solution. In such cases, evaluating the likelihood requires sampling the stochastic process or numerically solving its Fokker-Planck equation~\cite{gillespie92}. However, if the variation in the rotation potential $V$ is small compared to thermal energies, the potential can be neglected, which greatly simplifies the estimation. In particular, when $V$ is independent of the rotation angle, \eq{L stochastic} is an Ornstein-Uhlenbeck (O-U) process~\cite{uhlenbeck30}. In this case, the likelihood is readily computed. The examples considered in this paper have small potentials, allowing us to apply this simplification.

\section{Friction of Rotating Nested Nanotubes} \sectlabel{nanotubes}

The nested nanotube is a prototypical nanomechanical assembly. It has been demonstrated experimentally~\cite{cumings00} and its structural and operational behavior has been extensively discussed. Because of this importance, many molecular modeling studies have examined single- and multi-walled nanotubes, providing a considerable body of theory against which to compare our proposed technique for evaluating friction. This includes prior studies on the rotational friction of nested nanotubes~\cite{cook13}. For this reason, we use nested nanotubes as a test case for our method.

\begin{figure}
\centering  
\includegraphics[width=\widefigwidth]{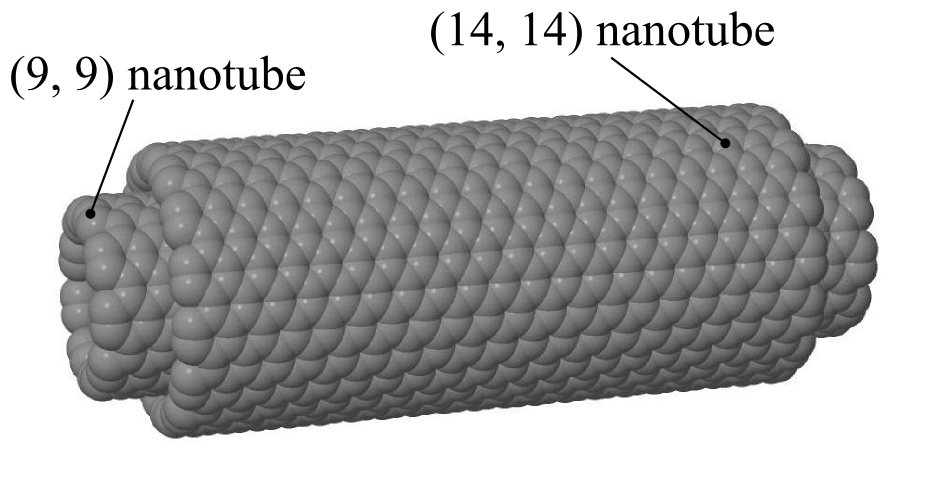}
\caption{Nested nanotubes. The inner tube is $6\,\nanometer$ long and the outer tube is $5\,\nanometer$ long. These consist of 882 and 1120 carbon atoms, respectively.
}
\figlabel{nanotube_overview}
\end{figure}

We applied the fluctuation-based method for evaluating friction to rotating nested nanotubes shown in \fig{nanotube_overview}. In this model system, a (9,9) single-wall carbon nanotube (SWCNT) rotates inside a (14,14) SWNCT. This example focuses on the rotary bearing itself, and ignores any damping that might arise from the housing to which the bearings would be connected in a molecular machine. 

The relatively small number of atoms in the nanotubes and their short damping time allow feasible simulations to show significant damping, even at relatively low rotation speeds, comparable to thermal speeds. This nanotube system thus provides a test case to compare damping estimated directly from the decrease in angular momentum with that estimated from fluctuations over a time scale much shorter than the damping time. Moreover, these simulations allow direct comparison with the extensive prior study of these nanotubes with simulations using other boundary conditions~\cite{cook13}.
\tbl{nanotube parameters} describes the inner nanotube, and \sectA{nanotube gromacs details} describes our simulations.

\begin{table}
\begin{center}
\begin{tabular}{lcc}
parameter    &	&   value \\ \hline
temperature	&$T$	& $300\,\Kelvin$\\
radius		&$r$		& $0.6\,\nanometer$\\
mass				&$m$	& $1.76 \times 10^{-23}\, \kilogram$\\
moment of inertia		&$J$		& $6.56 \times 10^{-42}\,\kgmsq$\\
thermal angular momentum	&$\Lthermal$	&  $1.65 \times 10^{-31}\,\kgmsqpers$\\
thermal angular velocity	&$\Wthermal$	&  $25.1\,\radian/\nanosecond$\\
\end{tabular}
\end{center}
\caption{Parameters for the inner nanotube. The root-mean-square thermal angular momentum and velocity are $\Lthermal = \sqrt{\BoltzmannConstant T J}$ and $\Wthermal=\Lthermal/J$, respectively, where $\BoltzmannConstant$ is Boltzmann's constant.}\tbllabel{nanotube parameters}
\end{table}

\subsection{Nanotube Behavior}

\begin{figure}
\centering  \includegraphics[width=\figwidth]{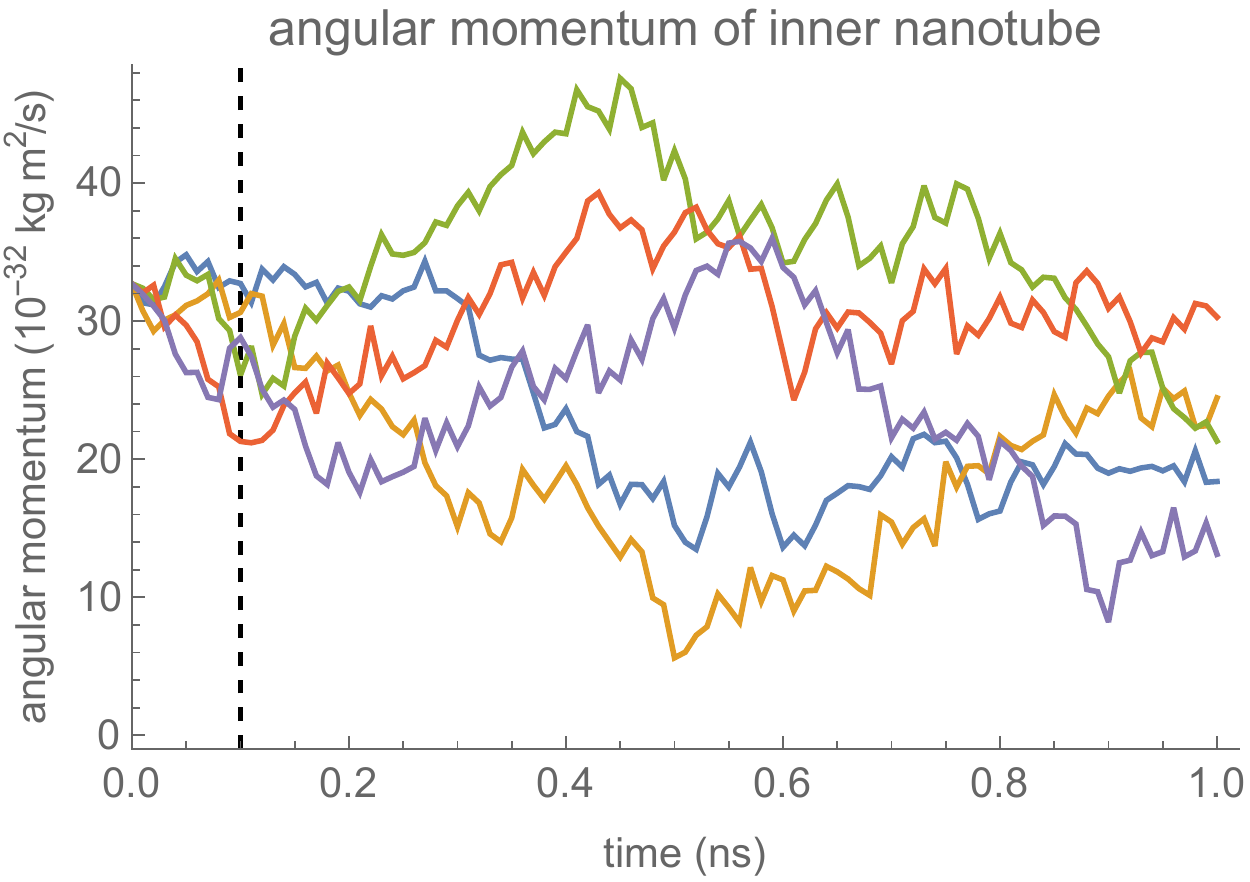}
\caption{The angular momentum of the inner nanotube about its axis vs.~time for each of the spin-down simulations, sampled at $0.01\,\nanosecond$ intervals. Successive samples are connected with straight lines, which do not show the high-frequency fluctuations between samples. The vertical dashed line indicates the extent of the truncated simulations used to estimate drag based on fluctuations.}
\figlabel{nanotube angular momentum}
\end{figure}

\fig{nanotube angular momentum} shows how the angular momentum of the inner nanotube changes during the simulations, starting with angular velocity $\omega_0=50\,\radian/\nanosecond$. The initial angular velocity is smaller, and closer to typical thermal velocities, than prior studies of this system~\cite{cook13}. Hence thermal fluctuations are relatively large, making it somewhat difficult to identify the decrease in average angular momentum even with the simulation time comparable to the damping time. This identification is even more difficult for the $0.1\,\nanosecond$ simulation time that we use for the fluctuation-based method.

\begin{figure}
\centering 
\includegraphics[width=\figwidth]{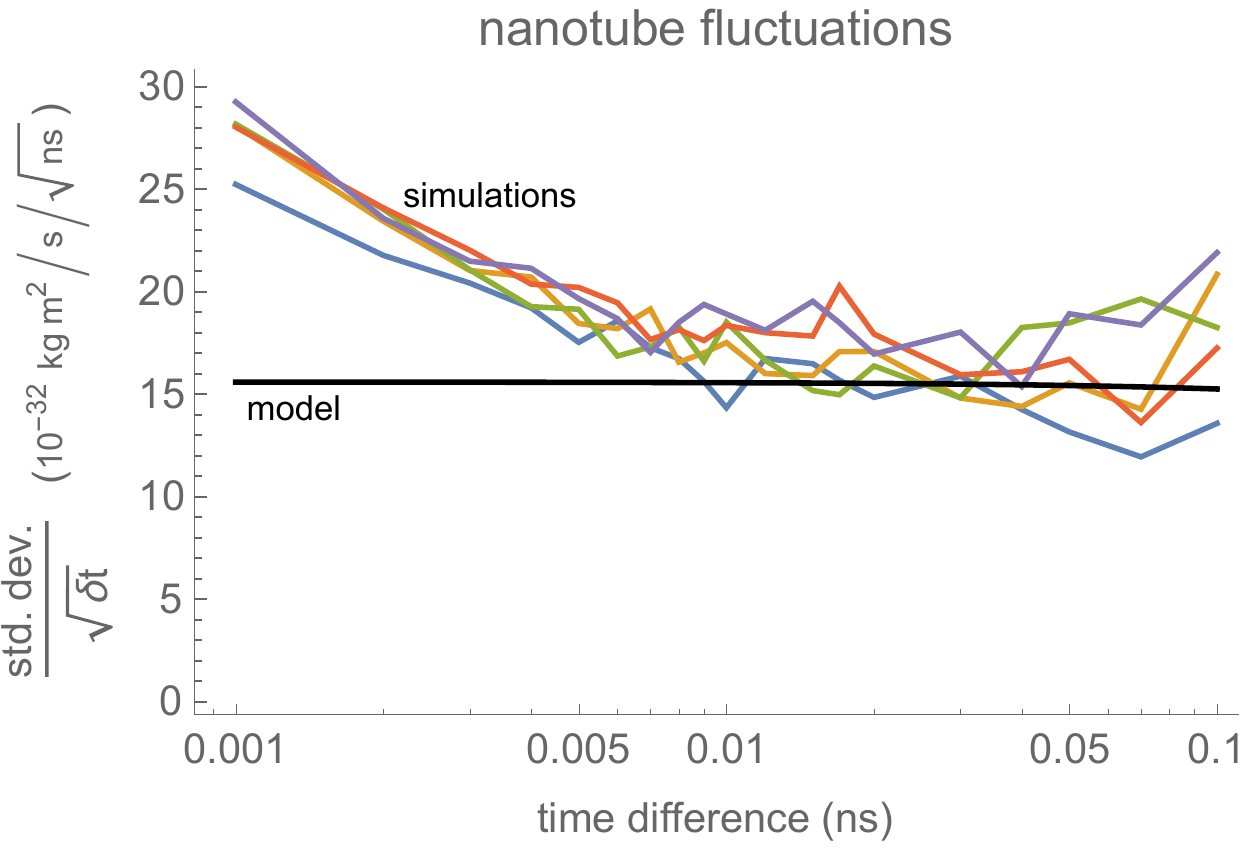}
\caption{Standard deviation of changes in nanotube angular momentum in time $\delta t$ divided by $\sqrt{\delta t}$, 
for each of the nanotube simulations (upper, colored curves). The solid horizontal black line shows the value of this ratio expected from the stochastic process model of the nanotubes. Times are shown on a logarithmic scale.}
\figlabel{nanotube fluctuations}
\end{figure}

To identify the time scale over which the nanotubes match the stochastic model of \eq{L stochastic}, \fig{nanotube fluctuations} shows a fluctuation plot, described in \sect{validity}. The curves are roughly the same for $\delta t$ above $0.01\,\nanosecond$. This indicates  \eq{L stochastic} approximates the nanotube motion for times longer than this. In particular, short simulations, of just $0.1\,\nanosecond$, should be more than enough to estimate damping using the fluctuation-based method.
These nested nanotubes have negligible potential barriers to rotation~\cite{servantie06}, so the potential is not relevant for friction estimation.

\subsection{Nanotube Drag}\sectlabel{nanotube drag}

\begin{table}
\begin{center}
\begin{tabular}{lcc}
method   	& simulation time	&   damping constant $\gamma$	\\ \hline
direct		& $1\,\nanosecond$	& $0.40(\pm0.3)/\nanosecond$\\ 
direct		& $0.1\,\nanosecond$	& $1.5(\pm1.1)/\nanosecond$\\ 
fluctuation		& $0.1\,\nanosecond$	& $0.45(\pm0.2)/\nanosecond$\\
\end{tabular}
\end{center}
\caption{Estimated damping constant, $\gamma$, for the nanotube using direct and fluctuation-based methods. The ranges given for the estimated parameters are approximate $95\%$ confidence intervals.}\tbllabel{nanotube drag parameters}
\end{table}

\tbl{nanotube drag parameters} compares the drag estimates from two methods: 
1) a direct estimation by fitting an exponential decay $e^{-\gamma t}$ to the average angular momentum of the simulations, and
2) the fluctuation-based method applied to the first $0.1\,\nanosecond$ of each simulation.
The damping estimates are similar, showing the fluctuation-based method provides a useful estimate using simulations much shorter than the damping time, which is $1/\gamma \approx 2.4\,\nanosecond$ in this case.
The corresponding drag is $\Kdrag = 2.9(\pm1.5) \times 10^{-33}\,\kgmsqpers$.
On the other hand, applying the direct method to the first $0.1\,\nanosecond$ of the simulations gives a poor estimate.

This estimate for nanotube drag is consistent with that of a previous study~\cite{cook13} based on simulations at higher speeds, above $25\,\gigahertz$. As described in \sectA{nanotube gromacs details}, this study gives $\Kdrag = 3.4(\pm0.8) \times 10^{-33}\,\kgmsqpers$. This consistency indicates the linear fit of frictional force vs.~speed of \cite{cook13} extends to the lower speed used in our simulation, i.e., $8\,\gigahertz$. 

We find similar behaviors when operating the nanotubes at higher speed and lower temperature, as described in \sectA{nanotube operating conditions}.


\section{Friction of an Acetylenic Single-Bond Rotor}\sectlabel{rotor}

Bearings consisting of single bonds are molecular rotary joints that avoid sliding surfaces. This contrasts with molecular rotors such as nested nanotubes, in which rotor and housing are not bonded and involve sliding friction from many atoms. 

This section describes and evaluates such a design. The rotary joints consist of covalent acetylenic bonds between rotor and housing. 
Their linear geometry makes acetylenic bonds ideal as low barrier linkages in rigid organic frameworks. This property has been exploited in nanoscale mechanical design, where several designs of the Nanocar~\cite{shirai05} employed acetylenic linkages as axles between fullerene wheels and a planar chassis.
The structure includes enough of a housing around the rotor to include long range interactions between the rotor and the rest of a molecular machine using the rotary joint. 
As with the nanotube bearing, we consider operation in vacuum at room temperature.

To evaluate rotor behavior, we used two computational techniques. First, we applied density functional theory (DFT) calculations to the rotor and part of the surrounding structure under symmetry constraints to determine how the rotor interacts with the housing.  This included the predicted rotational barrier due to the shortest range rotor-housing interactions, and the vibrational mode energies associated with the rotor motion.
Second, we used the DFT results to generate missing AMBER force field terms and RESP partial charges of the rotor and surrounding structure for the molecular dynamics simulations, which then were used to sample the rotor's motion within the full structure. We used these samples to estimate the drag.
\sectA{rotor gromacs details} describes our simulations.

\subsection{Rotor and Housing}

\begin{figure}
\centering  
\includegraphics[width=\widefigwidth]{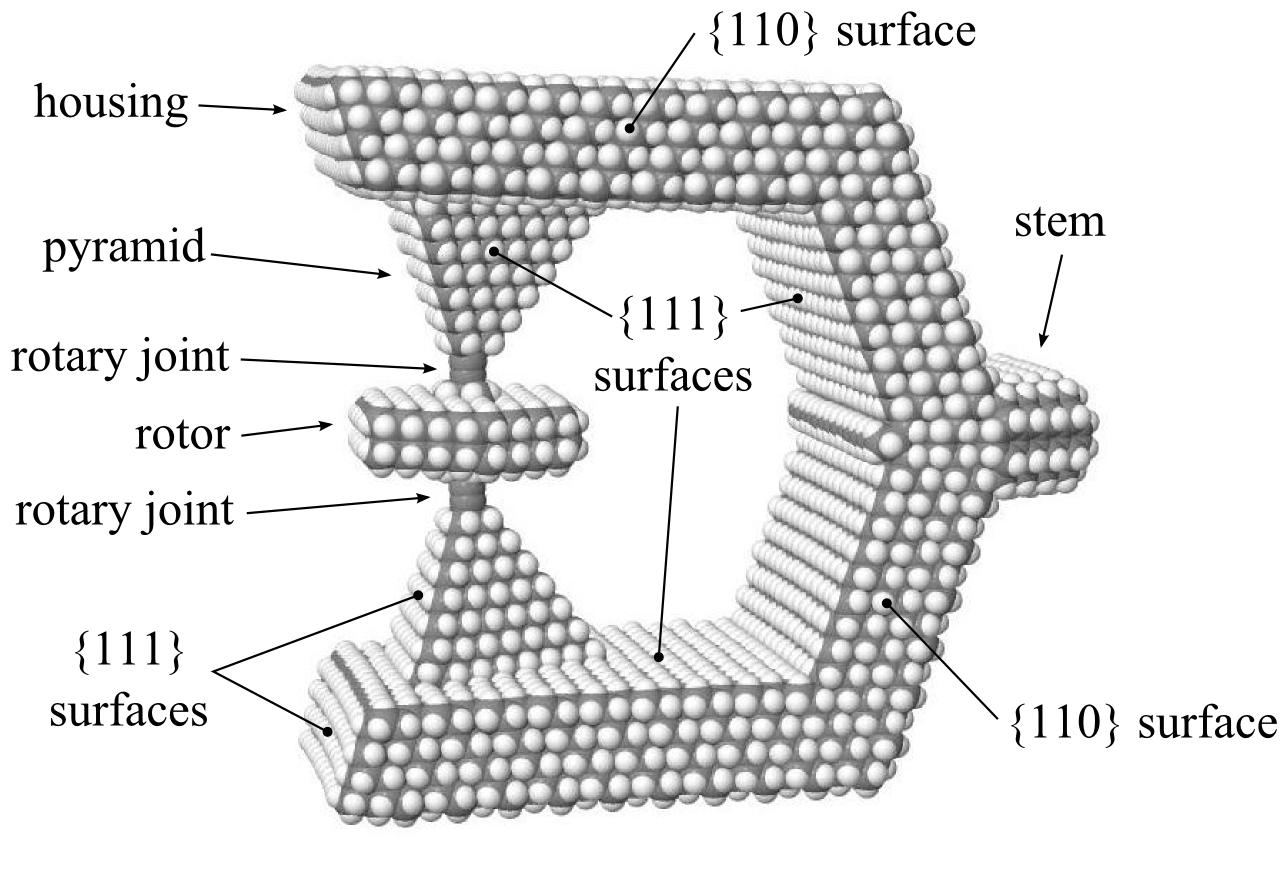}
\caption{A rotor bonded to a housing. This molecule contains 10345 atoms, 7693 carbon (gray) and 2652 hydrogen (white), and occupies a volume of about $4\,\nanometer \times 7\, \nanometer \times 7\, \nanometer$. The rotor consists of 200 carbon atoms and 150 hydrogens on its surface. The housing has 7493 carbon atoms and 2502 hydrogens.
The rotor is shown in the minimum energy ``staggered" configuration, which defines the origin of the rotation angle: $\theta = 0$.
}
\figlabel{rotor_overview}
\end{figure}

\fig{rotor_overview} shows our rotary system: a molecule supporting a mobile rotor with two coaxial molecular rotary joints attached to a stationary housing. 
The rotor is made of hexagonal diamond and suspended on two molecular rotary joints connected to a housing made of faceted cubic diamond. 
The rotor is bonded to the housing at each end by a so-called $C_2$ spacer, which resembles the organic molecule 2-butyne.
\tbl{rotor parameters} describes the rotor.
While it is not yet possible to synthesize this rotary joint molecule, \sectA{stability} describes why it would be chemically stable.

We designed this specific molecule as likely to give a rotary joint with low dissipation.
Specifically, this molecule
\begin{itemize}
\item provides a complete rotor and housing that includes all thermal motion in and around the rotor, and long-range interactions between the rotor and its housing

\item isolates the rotor motion from the housing vibrations by use of a stiff material (diamond)

\item has a low rotational energy barrier 

\item has closely matched rotor length and housing spacing so there is little strain (compression or stretching) between the $C_2$ spacers and either the rotor or the housing

\end{itemize}

This molecule has two properties that simplify computational studies. First, the high symmetry of the rotor and nearby parts of the housing reduces the computational cost of determining its properties, including geometries corresponding to extrema of the rotational potential, normal modes of vibration, and partial charges on the atoms.
Second, the molecule consists of just three atom types: sp3 (tetrahedral) carbon, terminal C-H bond hydrogen, and sp (acetylenic) carbon. These standard atom types readily transfer to any molecular force field for additional testing and validation.

\begin{table}
\begin{center}
\begin{tabular}{lcc}
parameter    &	&   value \\ \hline
temperature	&$T$	& $300\,\Kelvin$\\
mass				&$m$	& $4.24 \times 10^{-24}\, \kilogram$\\
moment of inertia		&$J$		& $1.76 \times 10^{-42}\,\kgmsq$\\
amplitude of potential	&$V_0$	& $4.15\times 10^{-22}\,\joule$\\
thermal angular momentum	&$\Lthermal$	&  $8.54 \times 10^{-32}\,\kgmsqpers$\\
thermal angular velocity	&$\Wthermal$	&  $48.5\,\radian/\nanosecond$\\
\end{tabular}
\end{center}
\caption{Rotor parameters. 
$V_0$ is the amplitude of the rotational potential given in \sect{potential}, and is equivalent to $0.25\,\kJmol$ and $60\,\calmol$. The root-mean-square thermal angular momentum and velocity are $\Lthermal = \sqrt{\BoltzmannConstant T J}$ and $\Wthermal=\Lthermal/J$, respectively, where $\BoltzmannConstant$ is Boltzmann's constant.}\tbllabel{rotor parameters}
\end{table}

\subsection{Forces and Rotation Potential}\sectlabel{potential}

Evaluating the rotor's behavior requires accurate forces between the rotor and the housing. 
A significant portion of these forces arises from partial charges in the rotor and housing, which we evaluated using the RESP partial charge model described in \sectA{partial charges}. 

We use these charges as part of the force field for our molecular dynamics simulations, both to sample rotor behavior for determining drag and to estimate the rotation potential $V(\theta)$. This potential is an aggregate property of the molecular dynamics force field. It quantifies the conservative torques on the rotor in the stochastic model given by \eq{L stochastic}.
As described in \sectA{potential details}, the potential energy as a function of rotation angle is well-approximated by
\begin{equation}\eqlabel{rotor potential}
V(\theta) = V_0 (1-\cos(3 \theta))
\end{equation}
where $V_0$ is the potential amplitude given in \tbl{rotor parameters}. The potential has the threefold symmetry of the rotor structure.

This potential barrier, $2V_0 = 8.3\times 10^{-22}\,\joule$, is comparable to estimates of barriers for rotation about triple bonds~\cite{kottas05}. 
This barrier is about $0.2 \BoltzmannConstant T$, which is relatively small compared to thermal energy.

The potential applies a torque $-dV/d\theta$ to the rotor. The maximum torque is $3V_0 = 1.2\,\piconewton\,\nanometer$.
As an estimate of corresponding forces, if the maximum torque is mostly at the rotor's outer rim, with radius $r=0.9\,\nanometer$ (see \fig{rotor_alone} in \sectA{stability}), the force is $3V_0/r = 1.4\,\piconewton$. If instead the torque mainly acts through the axle, of radius $r=0.05\,\nanometer$, the force is $25\,\piconewton$.

\subsection{Rotor Behavior}

\begin{figure}
\centering  \includegraphics[width=\figwidth]{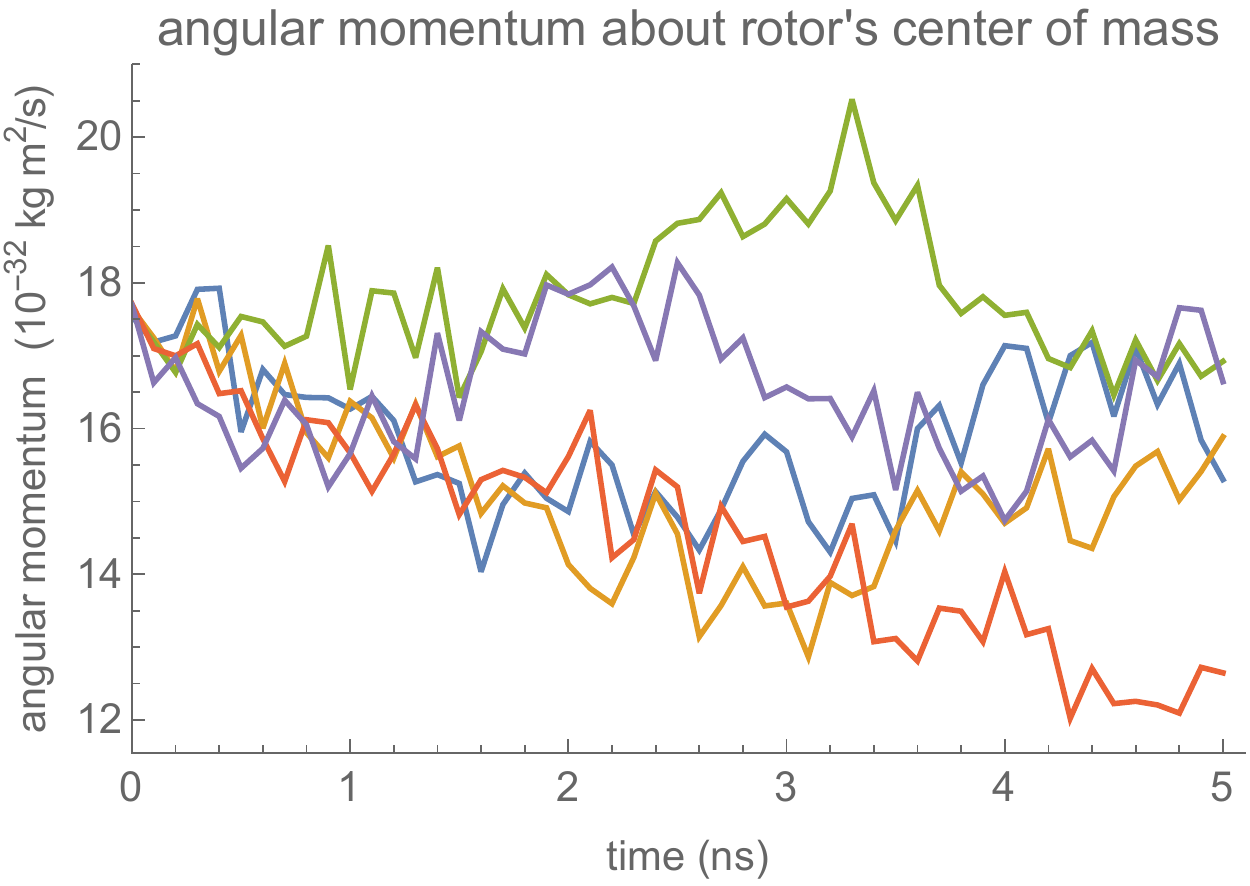} 
\caption{The $z$ component of rotor angular momentum about the center of mass vs.~time for each of the spin-down simulations, sampled at $0.1\,\nanosecond$ intervals. Successive samples are connected with straight lines, which do not show the high-frequency fluctuations between samples.}
\figlabel{rotor angular momentum}
\end{figure}

\fig{rotor angular momentum} shows the $z$ component of the rotor angular momentum about its center of mass during five spin-down MD simulations, starting with angular velocity $\omega_0=100\,\radian/\nanosecond$. During the $5\,\nanosecond$ simulation time, the angular momentum does not decrease significantly, indicating the damping time is considerably larger than the simulation time. 

The atoms of the rotor change position slightly during the simulations. However, due to the strong bonding, these motions have only a small effect on the moment of inertia. In particular, the standard deviation of the moment of inertia about the rotor axis, $J$, is $6 \times 10^{-45}\,\kgmsq$, less than $1\%$ of the mean value, given in \tbl{rotor parameters}.

\begin{figure}
\centering 
\includegraphics[width=\figwidth]{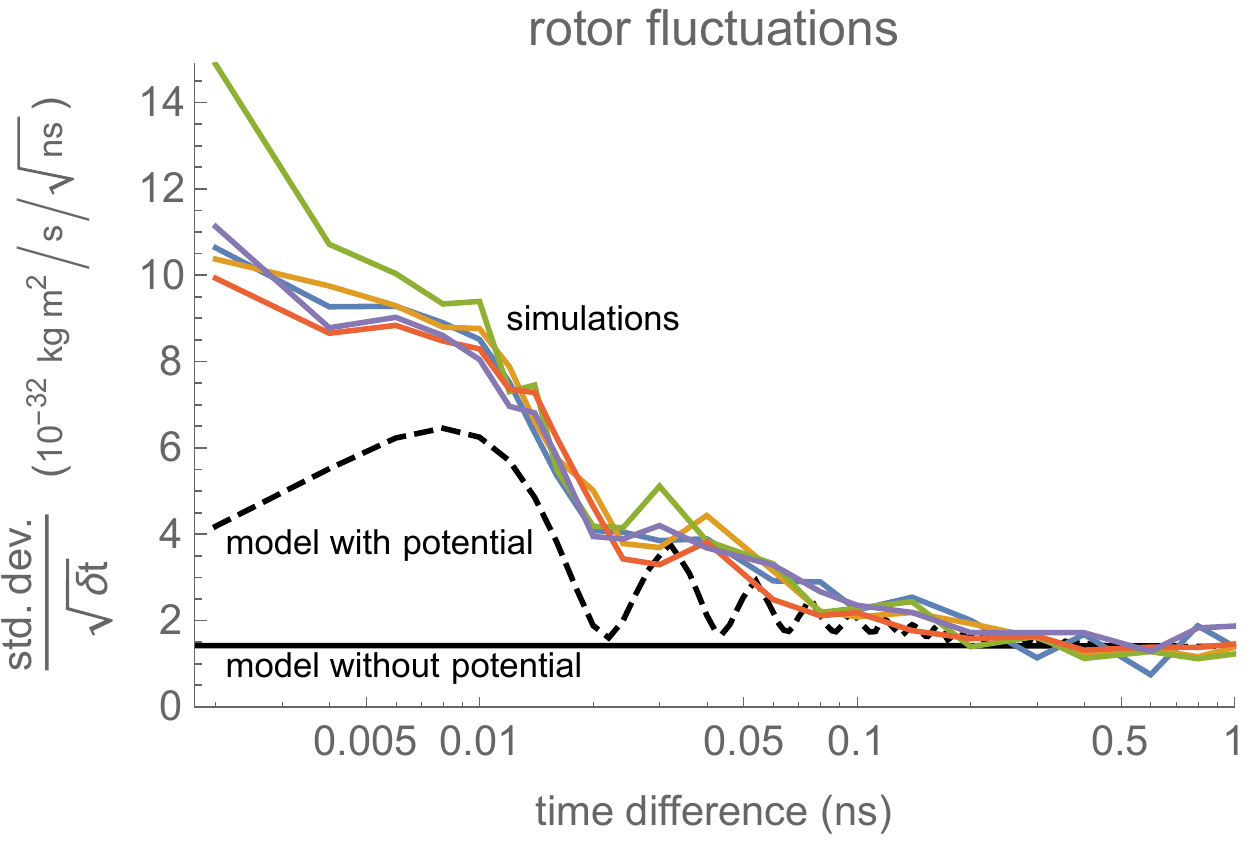}
\caption{Standard deviation of changes in rotor angular momentum in time $\delta t$ divided by $\sqrt{\delta t}$, 
for each of the rotor simulations (upper, colored curves). The dashed and solid black curves show the value of this ratio expected from the stochastic process model of the rotor  
   in \sect{rotor stochastic model}, with and without the potential, respectively. Times are shown on a logarithmic scale.}
\figlabel{rotor fluctuations}
\end{figure}

To identify the time scale over which the rotor matches the stochastic model of \eq{L stochastic}, \fig{rotor fluctuations} shows a fluctuation plot, described in \sect{validity}. The values are roughly the same for $\delta t$ above $0.2\,\nanosecond$ or so. For shorter times, the observed fluctuations are larger than the expected $\sqrt{\delta t}$ behavior from \eq{L stochastic}.
This indicates the stochastic model does not correctly describe short-time motion of the rotor, below about $0.2\,\nanosecond$ or above $5\,\gigahertz$. This conclusion is similar to that from the peaks in the power spectrum shown in \fig{power spectrum}.
These observations indicate  \eq{L stochastic} provides a reasonable approximation of the rotor's motion for times longer than about $0.2\,\nanosecond$.

The potential is relatively small compared to $\BoltzmannConstant T$. In addition, $\delta t$ above $0.2\,\nanosecond$ involves motion of more than a complete revolution of the rotor for the angular velocities in the simulation, which remain close to the initial $100\,\radian/\nanosecond$ because damping does not slow the rotor much during the $5\,\nanosecond$ simulation time. Thus the small effects of the potential are further reduced by the rotor averaging over the potential during the time steps we consider.

\subsection{Rotor Drag}\sectlabel{rotor drag}

\tbl{rotor drag parameters} gives the parameters maximizing the likelihood using increments with time step $\delta t=0.5\,\nanosecond$. Other choices of $\delta t$ in the range $0.2\mbox{--}1.0\,\nanosecond$ give the same values within the range of the confidence intervals. The estimated fluctuation magnitude, $\sigma$, is consistent with \fig{rotor fluctuations} for this range of $\delta t$.

\begin{table}
\begin{center}
\begin{tabular}{lcc}
parameter    &	&   value \\ \hline
damping constant		& $\gamma$	& $0.014(\pm0.005)/\nanosecond$\\
damping time			& $1/\gamma$	& $70(\pm40)\,\nanosecond$\\
fluctuation magnitude	& $\sigma$	& $1.4(\pm0.3) \times 10^{-32}\,\fluctuation$\\
frictional drag coefficient	& $\Kdrag$	& $2.4(\pm0.9) \times 10^{-35}\,\kgmsqpers$\\
\end{tabular}
\end{center}
\caption{Estimated drag parameters for the rotor. The ranges given for the estimated parameters are approximate $95\%$ confidence intervals.}\tbllabel{rotor drag parameters}
\end{table}

For example, with the angular velocity in our simulations, $\omega_0=100\,\radian/\nanosecond$, the time for a full rotation is $t=2\pi/\omega_0=0.06\,\nanosecond$. From \eq{dissipation}, the frictional energy dissipation during that rotation is $2\pi \Kdrag \omega_0\approx 1.5 \times 10^{-23}\,\joule$.
By comparison, a full rotation of this molecular rotor moves through three potential wells, each of depth $2V_0$ (see \tbl{rotor parameters}). Thus climbing the potential barriers requires $6V_0\approx 2.5\times10^{-21}\,\joule$. However, this work could, in principle, be recovered as the rotor moves into each potential well. 
The energy dissipated by friction is about 100 times smaller than the (possibly recoverable) energy required to climb the potential, which in turn is an order of magnitude smaller than the experimentally measured energy, $5\times 10^{-20}\,\joule$, to force rotation of a surface-bound molecule~\cite{loppacher03}.

\subsection{Stochastic Process for Rotor Behavior}\sectlabel{rotor stochastic model}

The stochastic process \eq{L stochastic} with parameters from \tbl{rotor drag parameters} gives a model of the rotor as a thermally-fluctuating molecular machine with a single degree of freedom. As a consistency check on this model and its range of validity, we compare the behavior predicted by this stochastic process with that from the simulations.

\fig{rotor fluctuations} shows one such comparison.
The rotational potential accounts for some of the increase in fluctuations at short times. In particular, the rotor converts between kinetic and potential energy as it moves between the top and bottom of the potential wells. For our simulations, the angular velocity remains close to its initial value $100\,\radian/\nanosecond$.  The potential, in \eq{rotor potential}, is 3-fold symmetric, so rotation from one potential well to the next requires rotation through $2\pi/3 \approx 2\,\radian$, which occurs in $0.02\,\nanosecond$, corresponding to $50\,\gigahertz$. This motion leads to the oscillating fluctuation size shown in the dashed curve of \fig{rotor fluctuations}.

\begin{figure}
\centering  \includegraphics[width=\figwidth]{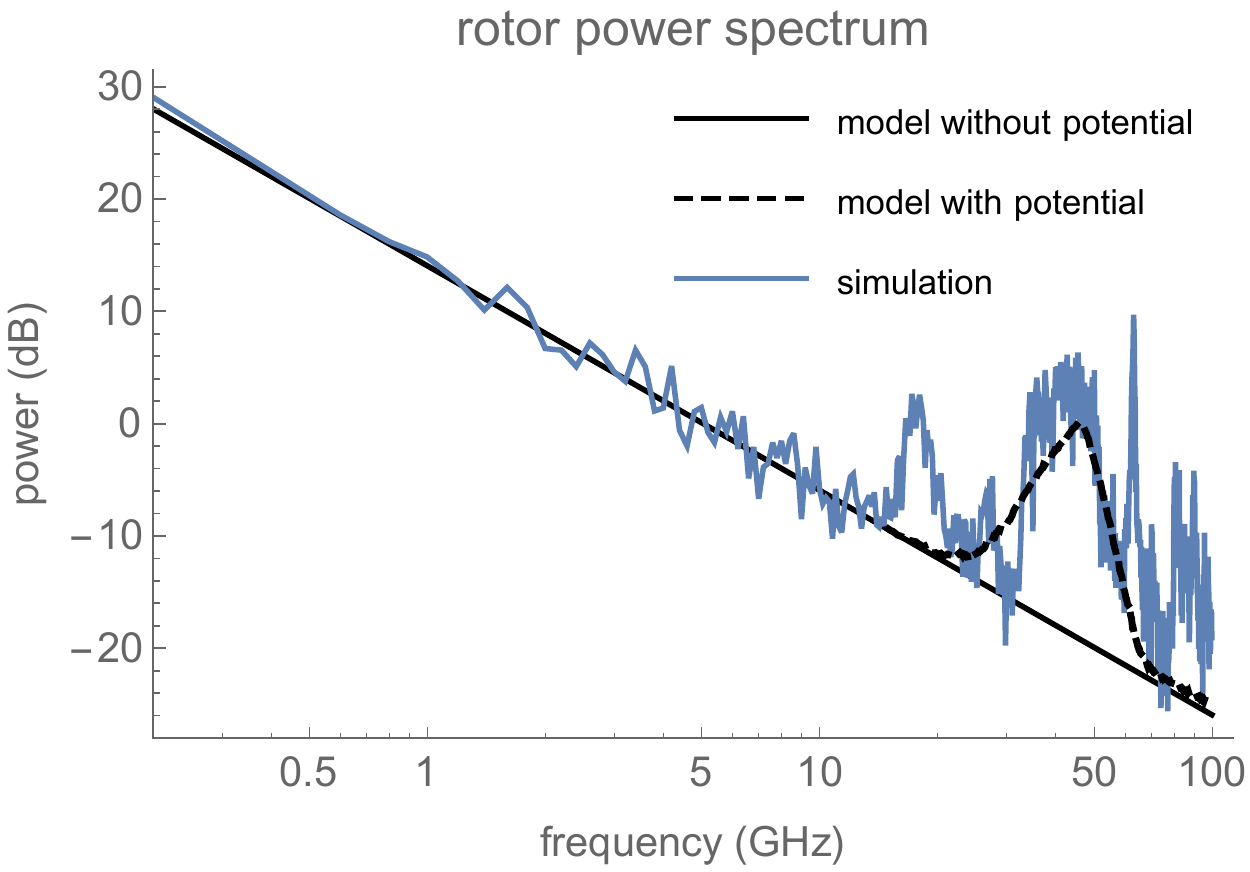}
\caption{Average of the angular momentum power spectra from the spin-down simulations. The horizontal axis shows frequency on a logarithmic scale. Power values are in decibels, normalized so the total power equals the sum of the squared angular momenta from the simulations. 
For comparison, the dashed and solid black curves show the power spectrum of the stochastic model of the rotor, \eq{L stochastic}, with and without the potential, respectively.
}
\figlabel{power spectrum}
\end{figure}

Another perspective on short-time rotor behavior is the average of the simulations' power spectra, shown in \fig{power spectrum}. 
By comparison, the power spectrum of an O-U random process is proportional to $1/(\gamma^2 + (2\pi f)^2)$~\cite{gillespie92}. For frequencies large compared with $\gamma$, this power decreases as $1/f^2$, and hence appears as a straight line on a log-log plot. \fig{power spectrum} compares this behavior with that of the rotor simulation, showing close correspondence with the rotor's angular momentum up to about $10\,\gigahertz$. 

The behavior including the potential (dashed curve in \fig{power spectrum}) has the broad peak near $50\,\gigahertz$, showing the potential ripple accounts for both the frequency (due to time it takes rotor to move over one potential barrier at $100\,\radian/\nanosecond$) and the magnitude of the peak.

Above about $50\,\gigahertz$, both the fluctuation plot and power spectrum show deviations from the stochastic process model, even when including the potential. One source of these deviations is temporal correlations in torques applied to the rotor, e.g., due to movement of its axis of rotation. The oscillation frequencies of the structure's normal modes indicate the time scale of these motions.
We evaluated the normal modes of the rotor-housing structure using GROMACS.
The lowest mode is the rotor twisting about its axis, at about $7.24\,\gigahertz$. 
This is close to the oscillation frequency of the potential minimum: $3\sqrt{V_0/J}/(2\pi) = 7.3\,\gigahertz$. 
This mode corresponds to the rotary motion given by the stochastic model.
The next lowest modes involve the rotor tilting, at around $100\,\gigahertz$. These modes characterize deviations from the stochastic model, and show the large separation in time scales between the rotary degree of freedom and other modes in the structure. This separation is the expected behavior for a rotary bearing consisting of strongly bonded structures and weak barriers to rotation.

These comparisons between simulations and the stochastic process with the estimated parameters show the atoms in the rotor and housing behave approximately as a stochastic rigid rotor, with a single degree of freedom around the rotary joint. The atoms' other degrees of freedom act as a source of uncorrelated random fluctuations, i.e., a thermal bath.

\section{Discussion}\sectlabel{discussion}

This paper applies a fluctuation-based method for evaluating friction in molecular machines to rotary bearings. 
The method performs well even when simulation times are short compared to the machine's damping time. Thus, the method can use much less computational time than a direct estimate from the average decrease in energy. 

The method was applied to nested nanotubes and a rotor covalently bonded to a housing by single acetylenic bonds.
For the nested nanotubes, damping time is $1/\gamma=2.4\,\nanosecond$ and drag coefficient is $\Kdrag = 2.9 \times 10^{-33}\,\kgmsqpers$ (\sect{nanotube drag}). The bonded rotor joint has $1/\gamma=70\,\nanosecond$ and $\Kdrag = 2.4 \times 10^{-35}\,\kgmsqpers$ (\sect{rotor drag}). 
From \eq{dissipation}, this means rotational friction dissipates $120$ times less energy for the bonded rotor than for the nested nanotubes when they rotate at the same speed.
Thus acetylenic rotary joints can have exceptionally low dissipation, which could make them useful bearings in future molecular machines.

Machine design often focuses on performance and ease of manufacture. For molecular machines that are not yet possible to fabricate, an important additional criterion is the computational cost to simulate their behavior. As an example of designs that simplify drag analysis, rotors with small potential barriers compared to thermal energy allow neglecting the potential in the stochastic model used with the fluctuation-based method. 
In addition, strong bonding among degrees of freedom other than those of the machine's operation (in this case the single rotational degree of freedom) allows treating the machine as a rigid body, with other degrees of freedom providing uncorrelated random fluctuations.

An open question is the range of conditions under which the fluctuation-based method performs well. For instance, potentials with larger barriers to rotation, less rigid molecular housings or less symmetric rotational geometry than the examples considered in this paper may significantly reduce the accuracy of modeling the rotor as a stochastic process with a single degree of freedom. 
Since neglecting the rotational potential simplifies the analysis, a specific issue for future work is determining design rules under which a molecular rotor is sufficiently well-approximated as having zero potential when evaluating its friction.

The fluctuation-based method discussed in this paper could evaluate drag in a range of operating conditions.
Examples are how rotary friction depends on rotor size, rotation speed and temperature. This could require accounting for additional physical effects. For instance, when operating well below room temperature, the variation in the potential is significant compared to $\BoltzmannConstant T$, leading to more complicated behavior than seen in the examples of this paper~\cite{kottas05,vacek01}.
Moreover, at low temperatures, quantum effects alter how energy spreads among degrees of freedom and introduce additional dissipative mechanisms, as described in \sectA{quantum}.

The interactions between rotor and housing arise mainly from forces between the rotor and nearby atoms in the pyramid connectors. These interactions largely determine the potential barrier to rotation. Future work could more precisely quantify these short-range interactions and the extent to which they are responsible for the friction. 
Identifying the main contributions to friction will suggest how changes to the housing structure affect dissipation, e.g., by altering stresses on bonds near the rotary joint.

The fluctuation-based method can apply to many molecular machines in addition to the rotary bearings considered in this paper. 
In particular, a single bearing will only be a small part of a useful molecular machine using the bearings, such as a computer~\cite{merkle16}. Designing such machines, containing large numbers of atoms, will require efficient computational methods.
Thus the fluctuation-based method could aid computational explorations of a wide variety of designs to identify low-power molecular machines. Fluctuations can estimate drag from simulations much shorter than damping time, which is particularly useful for low-friction molecular machines with many atoms. However, the simulation must still be long enough to average over dynamics of other modes of behavior than the degree(s) of freedom involved in the machine's intended operation. Otherwise fluctuations may not be independent, as assumed by the stochastic process model. More specifically, the method is useful when the correlation time of fluctuations is much smaller than the damping time of the machine's operation. Molecular machines made of strongly bonded structures can have such separation of time scales, as illustrated by the examples discussed in this paper. Thus the method described in this paper could be a useful tool for evaluating dissipation in such machines.

\section*{Acknowledgements}

We thank Jeremy Barton, Robert A. Freitas~Jr., Michael S. Marshall, Ralph C. Merkle, Thomas E. Moore and James Ryley for helpful discussions.
We appreciate Eugene Cook's clarification of Ref.~\cite{cook13}.

\newpage
\appendix

\begin{center}
{\huge \textbf{Appendices}}
\end{center}


\section{Molecular Dynamics Simulations}\sectlabel{gromacs details}

\sect{MD simulation} gives an overview of our molecular dynamics simulations, which used GROMACS (v.~5.0.4) \cite{van2005gromacs, hess08}.
The GROMACS simulations used double precision on a PC with an Intel Core i7-5960X $3.00\,\gigahertz$ processor and 48\,GB RAM, running on 64-bit Ubuntu~14.04.4 LTS. 
GROMACS structure, topology, configuration files and supporting scripts are available on request.

For both the nanotube and rotor simulations, the warmup stage starts with all atoms at zero velocity and couples them to a heat bath, i.e., the simulations used a canonical (NVT) ensemble. The spin-down simulations used a microcanonical (NVE) ensemble, i.e., without temperature control. 

The warmup and spin-down simulations are performed in vacuum and use $1\,\femtosecond$ time steps. In accordance with best-practices for molecular dynamics simulation in vacuum while reducing computational resource use, all GROMACS simulations were performed under periodic boundary conditions using $25\times25\times25\,\nanometer$ boxes with rlist, rcoulomb, and rvdw cut-offs of $10\,\nanometer$, particle mesh Ewald (PME) \cite{darden1993particle} coulombtype and vdwtype (with pme-order of 6), and the Verlet cutoff scheme under a grid-based neighbor list determination.

The box size used for periodic boundary conditions is larger than the sum of half the molecule's dimensions and the cut-off lengths. The volume of this box remains constant during the simulation. Since the molecule is in vacuum, the simulation has both constant volume for the full simulation volume and zero pressure on the molecule.

The warmup simulation uses the following parameters: 50,000 steps at $1\,\femtosecond/\mbox{step}$ ($50\,\picosecond$ total simulation time), xyz periodic boundary conditions with Verlet cutoff, $10\,\nanometer$ cutoff distance (which exceeds the largest atom-to-atom distance in the structure, and is less than the distance between copies of the structure due to periodic boundary conditions), v-rescale temperature control with $1\,\picosecond$ time constant, and random seed set to 42, 43, 44, 45, or 46. These seeds give five different samples of the structure at $300\,\Kelvin$.
During the warmup simulations, the structures reach $300\,\Kelvin$ within about $10\,\picosecond$, indicating the $50\,\picosecond$ warmup simulation time is more than sufficient to eliminate transients during warmup.

\subsection{Molecular Dynamics for Nested Nanotubes}\sectlabel{nanotube gromacs details}

For our evaluation of nanotube drag, we focus on the inner nanotube and hold fixed the atoms on one edge of the outer tube. After the warmup, velocities added to the atoms of the inner tube give it initial angular velocity $\omega_0=50\,\radian/\,\nanosecond$, corresponding to $8\,\gigahertz$ rotation frequency. The subsequent spin-down simulations were for $1\,\nanosecond$. We performed five separate tests of the nanotubes. A 50-picosecond warmup nanotube simulation takes one minute to complete, while a five-nanosecond spin-down simulation takes about 20 minutes.

\tbl{nanotube_forcefield} provides the force field parameters~\cite{abrahamgromacs} used for the nested nanotube bearing simulations. 
The bond stretch and bond angle terms are based on the AMBER force field~\cite{wang2000well}. The Ryckaert-Bellemans dihedral terms are based on the bending rigidity of graphene~\cite{lu2009elastic}, and the Lennard-Jones terms are based on~\cite{servantie06b}.

\begin{table}
\begin{center}
\begin{tabular}{lcrl}
Interaction Type    & Parameter &  Value & Unit  \\ 
\hline
Morse Potential Bond Stretch    & $b$           &   0.142    & nm \\
Morse Potential Bond Stretch    &  $D$         &  722.7      & kJ/mol \\
Morse Potential Bond Stretch    &  $\beta$   &   18.33      & nm$^{-1}$ \\
Harmonic Angle Potential           &    $k$   	&   527.2 &  kJ/mol/rad$^2$         \\
Harmonic Angle Potential           & $\theta_0$  &  120.0  &  degrees \\
Ryckaert-Bellemans Dihedral     &   $C_1, C_3, C_4, C_5$ &  0.0  &  kJ/mol \\
Ryckaert-Bellemans Dihedral     &  $C_0$  & 17.843   &  kJ/mol \\
Ryckaert-Bellemans Dihedral     &  $C_2$  &  -17.843  &  kJ/mol \\
Lennard-Jones                           &  $\epsilon$ & 0.28454   &  kJ/mol \\
Lennard-Jones                           &  $\sigma$ &  0.341  &  nm \\
\end{tabular}
\end{center}
\caption{Force field parameters used for GROMACS nested nanotube bearing simulations.}
\tbllabel{nanotube_forcefield}
\end{table}

Our nanotube simulations are similar to the ``coast-down'' method of \cite{cook13}. Minor differences arise from our use of GROMACS instead of LAMMPS, the force field of \tbl{nanotube_forcefield} instead of AIREBO, and having the outer tube anchored at one edge instead of freely counter-rotating.

Our friction estimate is consistent with the linear fit to normalized values given in \cite{cook13}. This fit is $F^* = A v^*$ where $F^*$ and $v^*$ are normalized friction force and sliding velocities~\cite[Eq.~23 and 24]{cook13}, $A=4.62\times 10^{-4}$ and the fit is consistent with having zero friction at zero velocity.\footnote{The corrected linear relation between force and speed~\cite{cook16} corresponds to \cite[Fig.~9]{cook13} rather than \cite[Eq.~28]{cook13}.} Converting the friction force to a torque, and the sliding velocity to the corresponding angular velocity, the linear relation between $F^*$ and $v^*$ gives the rotational frictional drag coefficient described in \sect{stochastic model}:
\begin{equation}
\Kdrag = A \frac{N \BoltzmannConstant T r}{2\pi v_{\mbox{\scriptsize LJ}}}
\end{equation}
where $N=2006$ is the number of atoms in the nanotubes, $r=0.77\,\nanometer$ is the average radius of the nested tubes and $v_{\mbox{\scriptsize LJ}}=0.139\,\nanometer/\picosecond$ is the normalizing velocity~\cite[Eq.~24]{cook13}. Thus $\Kdrag = 3.4(\pm0.8) \times 10^{-33}\,\kgmsqpers$, with the range indicating the $95\%$ confidence interval, obtained from the corresponding interval for $A$~\cite{cook16}.

\subsection{Nanotube Damping and Operating Conditions}\sectlabel{nanotube operating conditions}

\tbl{nanotube drag parameters} gives drag estimates for one operating condition for the nanotubes. This appendix describes the drag for other operating conditions to assess our method's generality.

\begin{figure}
\centering \includegraphics[width=\figwidth]{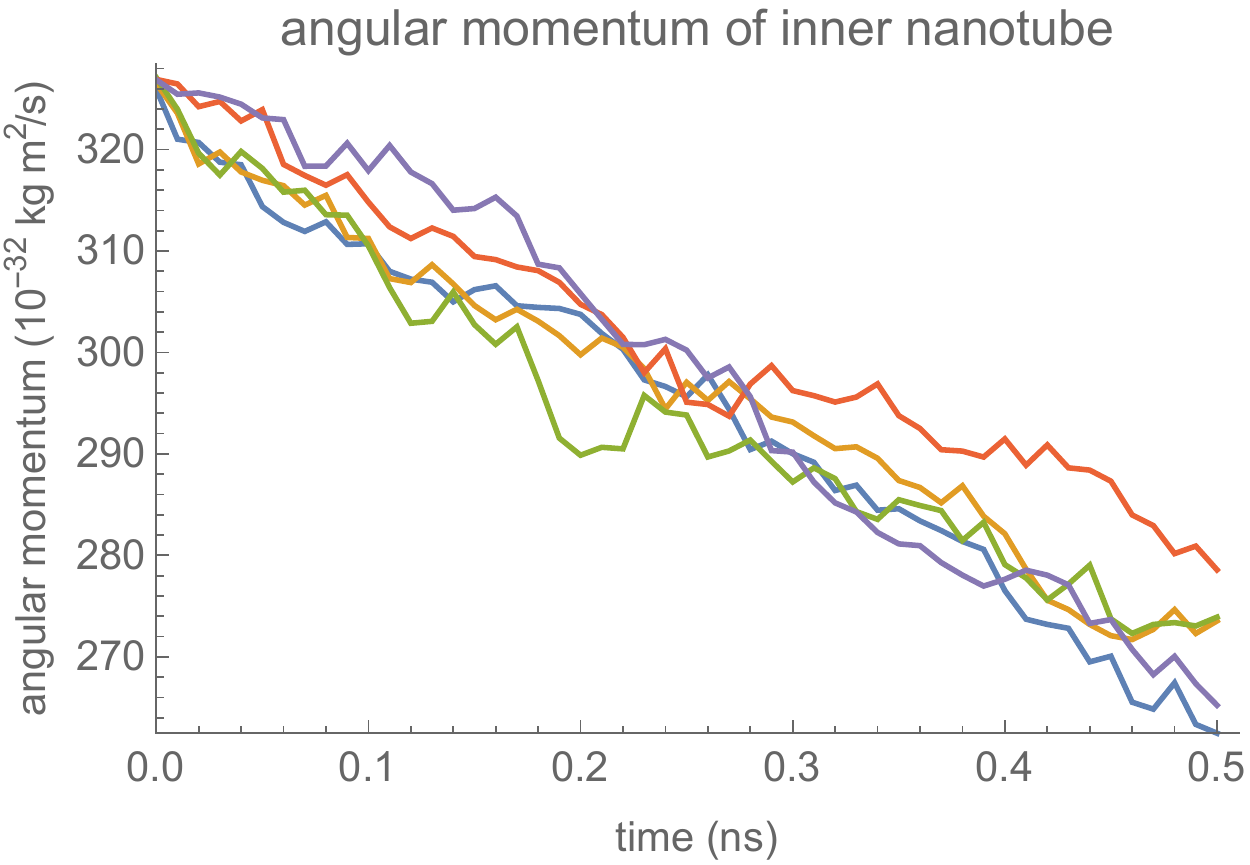}
\caption{The angular momentum of the inner nanotube about its axis vs.~time for spin-down simulations with initial angular velocity $\omega_0=500\,\radian/\nanosecond$, sampled at $0.01\,\nanosecond$ intervals. Successive samples are connected with straight lines, which do not show the high-frequency fluctuations between samples.}
\figlabel{nanotube angular momentum 500}
\end{figure}

\begin{table}
\begin{center}
\begin{tabular}{lcc}
method   	& simulation time	&   damping constant $\gamma$	\\ \hline
direct		& $0.5\,\nanosecond$	& $0.38(\pm0.03)/\nanosecond$\\ 
direct		& $0.1\,\nanosecond$	& $0.46(\pm0.1)/\nanosecond$\\ 
fluctuation		& $0.1\,\nanosecond$	& $0.41(\pm0.1)/\nanosecond$\\
\end{tabular}
\end{center}
\caption{Estimated damping constant, $\gamma$, for the nanotube using direct and fluctuation-based methods for 5 simulations with initial angular velocity $\omega_0=500\,\radian/\nanosecond$. The ranges given for the estimated parameters are approximate $95\%$ confidence intervals.}\tbllabel{nanotube drag parameters - fast}
\end{table}

When the nanotube rotates at speeds well above thermal velocity, fluctuations are relatively unimportant. In that case, the decrease in angular momentum is readily apparent, as illustrated in \fig{nanotube angular momentum 500} for nested nanotubes with initial angular velocity $500\,\radian/\nanosecond$. This initial speed is ten times the angular velocity considered in \sect{nanotubes}, and is well above typical thermal speeds (see \tbl{nanotube parameters}). \tbl{nanotube drag parameters - fast} compares drag estimated with the direct and fluctuation-based methods for this case. Because thermal fluctuations are relatively small, the direct method gives good estimates even over short times. Both methods give comparable drag estimates, and are consistent with the values at the lower speed, shown in \tbl{nanotube drag parameters}. This similarity is consistent with the linear relation between frictional torque and angular velocity found in other studies~\cite{cook13,servantie06b}, as well as the linear form for the drag in \eq{L stochastic}.

\begin{table}
\begin{center}
\begin{tabular}{lcc}
method   	& initial angular velocity	&   damping constant $\gamma$	\\ \hline
direct		& $500\,\radian/\nanosecond$	& $0.067(\pm0.01)/\nanosecond$\\ 
fluctuation		& $500\,\radian/\nanosecond$	& $0.066(\pm0.01)/\nanosecond$\\
direct		& $50\,\radian/\nanosecond$	& $0.071(\pm0.06)/\nanosecond$\\ 
fluctuation		& $50\,\radian/\nanosecond$	& $0.068(\pm0.03)/\nanosecond$\\
\end{tabular}
\end{center}
\caption{Estimated damping constant, $\gamma$, for the nanotube using direct and fluctuation-based methods at $80\,\Kelvin$ and two initial angular velocities, using $0.5\,\nanosecond$ simulation times. The ranges given for the estimated parameters are approximate $95\%$ confidence intervals.}\tbllabel{nanotube drag parameters - 80K}
\end{table}

Fluctuations and damping depend on temperature. 
As an example, \tbl{nanotube drag parameters - 80K}  shows the damping estimates from the two methods at $T=80\,\Kelvin$. We performed 5 simulations for each of the two initial angular velocities shown in the table. In this case, the thermal angular velocity is $\Wthermal=13.0\,\radian/\nanosecond$ so even the smaller initial speed, $\omega_0=50\,\radian/\nanosecond$, is well above the thermal speed. This means fluctuations are less important than at higher temperatures, and both the direct and fluctuation-based methods give similar results, as shown in \tbl{nanotube drag parameters - 80K}. At $80\,\Kelvin$, the damping time $1/\gamma$ is about $15\,\nanosecond$, so there is little damping during our $0.5\,\nanosecond$ simulations.

The fluctuation-dissipation theorem, \eq{fluctuation dissipation}, relates the fluctuation ($\sigma$) and damping ($\gamma$) parameters to temperature ($T$). Specifically the ratio $\sigma^2/\gamma$ is proportional to $T$. However, this relation does not determine how the two parameters vary individually. We identify this dependence by fitting the parameter estimates from our simulations at $80\,\Kelvin$ and $300\,\Kelvin$ to power-laws, $\sigma \propto T^\alpha$ and $\gamma \propto T^\beta$, with $2\alpha-\beta=1$ to match \eq{fluctuation dissipation}. This gives $\alpha = 1.23\pm0.08$ and $\beta = 1.47\pm0.15$.
This is consistent with a study of nanotube friction over a wider range of temperatures that found $\beta = 1.53\pm0.04$~\cite{servantie06b}.

\subsection{Molecular Dynamics for the Bonded Rotary Joint}\sectlabel{rotor gromacs details}

The boundary condition for the simulation holds fixed the atoms in the stem indicated in \fig{rotor_overview}. After warmup, we set the rotor's initial angular velocity to $\omega_0=100\,\radian/\,\nanosecond$, corresponding to rotation at $16\,\gigahertz$. The subsequent spin-down simulations were for $5\,\nanosecond$. We performed five rotor simulations. A 50-picosecond warmup rotor simulation takes less than an hour to complete, while a five-nanosecond simulation of rotor spin-down takes about 44 hours. 

Structure optimizations, normal mode analyses, and molecular dynamics simulations 
used the AMBER99 \cite{wang2000well} force field C-H and C-C bend, stretch, dihedral, and C6/C12 parameters serving as a basis for the calculations. Terms for the C\#C (triple bond) and C-C\# bonds were defined as given in Ref. \cite{wang2004development}.

The full structure (rotor plus housing) is first energy-minimized without applying anchors to any part of the structure. Next, anchors are applied to 514 atoms in the stem of the housing, and the structure is heated to $300\,\Kelvin$ in a warmup simulation.

The output of each of the five warmup simulations is a molecular structure in G96 format which contains the position and velocity for each atom in the structure. The initial angular velocity of the rotor is set by using a Matlab/GNU~Octave script to adjust the velocities of individual atoms. First the principal axes of the rotor are determined, and then the full system is rotated into a coordinate frame in which the $z$ axis aligns with the rotor's main principal axis. The component of the initial angular momentum due to thermal excitation about the main principal axis $L_z^{\mbox{\scriptsize warmup}}$ and initial polar moment of inertia about the main principal axis $J_z$ are calculated. A correction velocity of
\begin{equation}
\vec{v}_{corr} = \left[ \begin{array}{c} 0 \\ 0 \\ \omega_0 - L_z^{\mbox{\scriptsize warmup}}/J_z  \end{array} \right] \times \left[ \begin{array}{c} r_{xi} \\ r_{yi} \\ 0 \end{array} \right]
\end{equation}
is added to the $i$th atom in the rotor, where $r_{xi}$ and $r_{yi}$ are the $x$ and $y$ distances, respectively, from atom $i$ to the main principal axis. The full system is then rotated back into its original coordinate frame. This sets the rotational speed of the rotor to $\omega_0$, while leaving the initial angular velocities about the other two axes unchanged. 

The spin-down simulation uses the speed-adjusted G96 files as inputs and anchors the atoms in the stem. 
The simulation parameters are: 5,000,000 steps at $1\,\femtosecond/\mbox{step}$ ($5\,\nanosecond$ total simulation time), xyz periodic boundary conditions with Verlet cutoff, $10\,\nanometer$ cutoff distance.

\subsection{Energy and Temperature Drift for the Bonded Rotary Joint}

\begin{table}
\begin{center}
\begin{tabular}{lcccc}
Quantity    &	Average &  Standard Deviation & Total Drift &  Units\\ 
\hline
Total Energy        &      $1.7\times 10^5$  &   12  &  3.3     & \kJmol \\
Temperature        &     303  &   1.8 & -0.01 & \Kelvin        \\
\end{tabular}
\end{center}
\caption{Drift in energy and temperature for the rotor and housing during a GROMACS spin-down simulation. The total drift is the change between the start and end of the simulation.}
\tbllabel{drift}
\end{table}

The classical Dulong-Petit~\cite{ashcroft76} law gives a fair approximation for the heat capacity of the molecular machine at $300\,\Kelvin$, although the heat capacity of bulk diamond at this temperature is only one-fourth this value. 
The full system has $N = 10345 - 514 = 9831$ atoms that are free to move (514 atoms in the stem are frozen during the simulations). The total heat capacity of the system is then $3 N \BoltzmannConstant = 4.1 \times 10^{-19}\,\joule/\Kelvin$. If all of the initial kinetic energy of the rotor is dissipated as heat, the expected temperature increase would be $\Delta T = (J \omega_0^2/2) / (3 N \BoltzmannConstant) = 0.022 \,\Kelvin$ for an initial angular velocity of $\omega_0 = 100\,\radian/\nanosecond$. This upper bound on temperature increase is comparable to temperature drift due to numerical effects. Moreover, the five-nanosecond spin-down duration is too short to fully dissipate the rotor's kinetic energy, so the expected temperature increase is even smaller than this estimate. Thus there are no significant heating effects during the spin-down simulation.

\tbl{drift} shows the energy drift for the $5\,\nanosecond$ spin-down simulation with random seed = 42, calculated with the GROMACS utility  \code{g\_energy\_d}~\cite{abrahamgromacs}. The other simulations have similar values.
For the $N=9831$ atoms free to move, $3 N \BoltzmannConstant T = 1.2 \times 10^{-16}\,\joule$ (or $7.4\times 10^4\,\kJmol$), about half the average energy in the table. This is due to GROMACS including an offset of about $9.46\times  10^4\,\kJmol$ from the minimum-potential configuration of the rotor.

During each of the five spin-down simulations, the total system energy remained nearly constant. Specifically, over the full five nanosecond spin-down, the largest increase in total system energy, as computed by the GROMACS analysis tool \code{g\_energy\_d}, of any of the simulations was $6.51 \times 10^{-21} \joule$ ($4\,\kJmol$). This drift is small enough to not effect the conclusions of this study.

\subsection{Rotation Potential for the Bonded Rotary Joint}\sectlabel{potential details}

\begin{figure}
\centering  
\includegraphics[width=\figwidth]{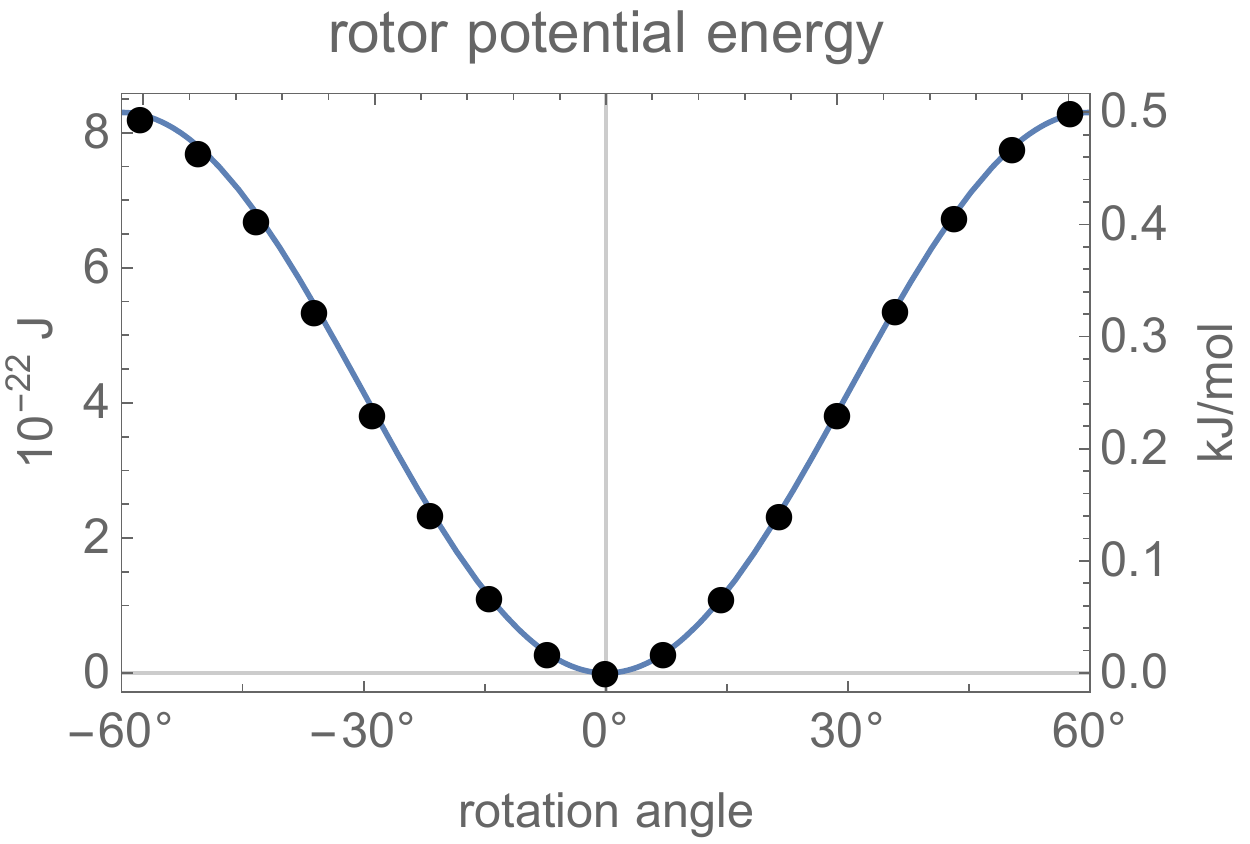}
\caption{Potential energy of the rotor and housing structure as a function of rotor rotation angle within one of the three symmetric potential wells. The points are the values measured with GROMACS and the curve is \eq{rotor potential}. The angle $0^\circ$ is the configuration shown in \fig{rotor_overview}.}
\figlabel{rotor_potential}
\end{figure}

\begin{figure}
\centering
\includegraphics[width=\widefigwidth]{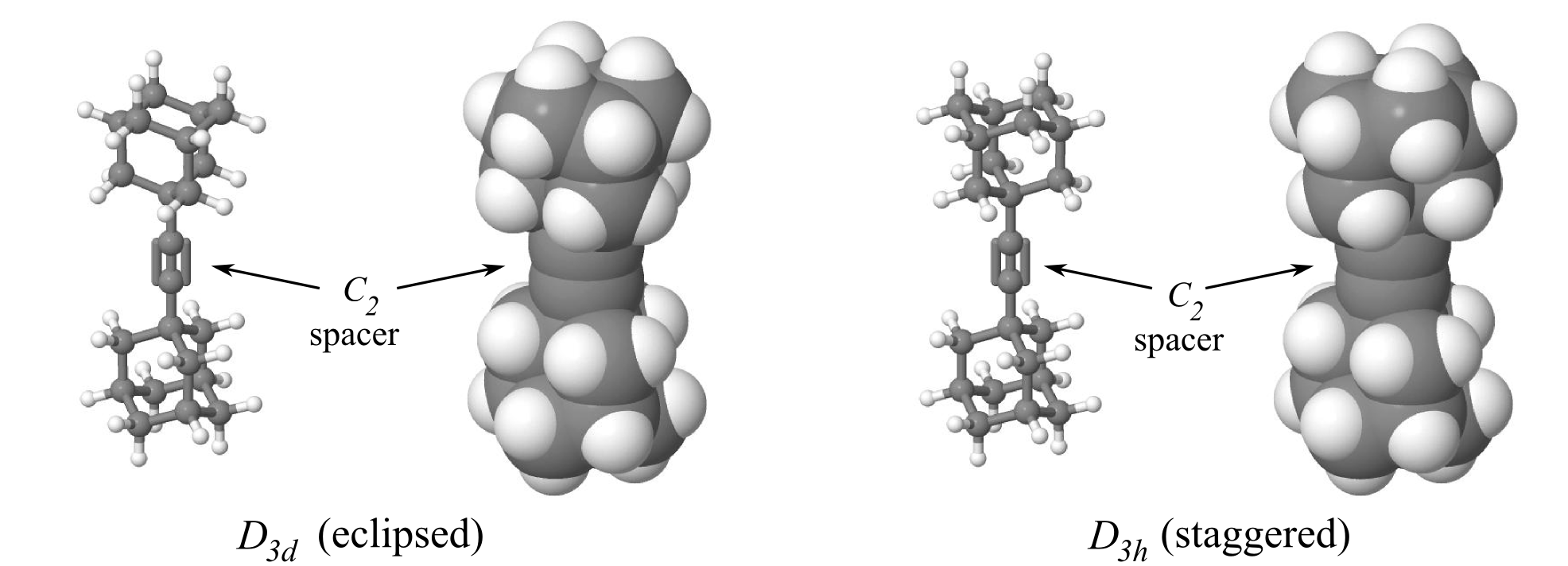}
\caption{Ball-and-stick and space-filling models of the molecular joint in two configurations: eclipsed and staggered on left and right, respectively.  $C_2$ spacers are indicated by the arrows. 
These geometries have molecular $D_{3d}$ and $D_{3h}$ symmetries~\cite{laidler82}, respectively.
}
\figlabel{staggered_eclipsed}
\end{figure}

The 3-fold symmetry of the rotor leads to a corresponding symmetry in the rotation potential: $V(\theta+120^\circ)=V(\theta)$. Using GROMACS, we measured the potential energy at various rotation angles in static configurations, i.e., at zero temperature. That is, for several angles $\theta$ between $-60^\circ$ and $60^\circ$, we evaluated the potential energy of each rotor atom according to the force field. The value of $V(\theta)$ is the sum of these individual potentials. \fig{rotor_potential} shows these measured values are very close to \eq{rotor potential}. 

For the purposes of our analysis, the key quantity is the height of the potential barrier $2V_0$ compared to $\BoltzmannConstant T$. As a check on the size of the potential, we performed a spin test at zero temperature. This test started the rotor spinning and measured the change in angular momentum as the rotor moved from one potential well to the next. This case has no thermal motion so the rotor is affected only by the potential. The observed changes were consistent with \eq{rotor potential} and the value $V_0$ in \tbl{rotor parameters}.

\fig{staggered_eclipsed} shows the configuration of atoms at the rotary joint corresponding to the extremes of the potential. The staggered geometry is the minimum of the rotational potential, while the eclipsed geometry is the maximum. The staggered configuration corresponds to rotor rotation angles (see \fig{rotor_overview}) of $\theta = 0, 120, 240$~degrees. The eclipsed configuration corresponds to angles of $\theta = 60, 180, 300$~degrees.

The potential arises mainly from non-bonding interactions between atoms of the rotor and the pyramid portions of the housing. 
The single $C_2$ spacer is, then, a compromise between minimizing the rotational barrier and maintaining stiffness along the rotation axis.

\subsection{Quantum Effects}\sectlabel{quantum}

Molecular dynamics simulations treat atoms as classical particles with defined bonds and persistent connectivity. The rotary molecular machines considered in this paper do not form or break bonds, thereby maintaining atom connectivity throughout the simulation. This appendix considers the validity of the classical approximation to the atoms' motions.

The thermal de~Broglie wavelength of an object with mass $m$ is $h/\sqrt{2\pi m \BoltzmannConstant T}$ which is $2\times10^{-3}\,\nanometer$ for the rotor, much smaller than its size. Thus the rotor is well-approximated as a localized classical mass.

At the temperature considered here, typical angular momenta of the nanotube and rotor (see \tbl{nanotube parameters} and \tbl{rotor parameters}) are about 1000 times larger than the quantum of angular momentum $\hbar=10^{-34}\,\kgmsqpers$. 
Thus the classical approximation of continuous variation of angular momentum is reasonable.
These observations indicate the rotor behaves as a classical object for operation at up to at least a few GHz.

Drag arises from spreading organized energy in rotor motion to the many other degrees of freedom of the molecular structure. Thus in addition to classical physics being sufficient to describe the rotor motion itself, molecular dynamics simulations assume classical behavior for other modes of motion. In particular, classical equipartition of energy among these modes only applies when thermal energy, $\BoltzmannConstant T$, is large compared to the quantum level spacing 
$h f$, where $f$ is a mode's frequency. 

The rotor speeds we consider (below $100\,\gigahertz$) have $\BoltzmannConstant T$ large compared to the quantum level spacing. Thus the classical behavior assumed in molecular dynamics simulations is a reasonable approximation for the rotor itself.

In addition to the rotary motion, the simulations include much higher frequency modes, particularly the hydrogen bond motions. Resolving these motions requires simulation time steps, $1\,\femtosecond$, much shorter than the time scales of the rotary behavior. For these high-speed motions, the quantum energy spacing is large compared to $\BoltzmannConstant T$. Thus these modes will have less energy than predicted by classical equipartition and GROMACS simulations. As a check on whether this suppression affects the rotor drag results, we repeated the simulations with the GROMAS H-bond constraint, which effectively keeps the hydrogen-carbon bonds in their ground state throughout the simulation. With this constraint, we find the same value for the rotor drag, within the reported confidence intervals, as in our original simulations. Nevertheless, this constraint significantly reduced the short-time fluctuations. With this reduction, the fluctuations closely match those expected from the stochastic process model (shown as the dashed curve in \fig{rotor fluctuations}) for time differences as short as $1\,\picosecond$.
This correspondence suggests the large fluctuations at short times seen in \fig{rotor fluctuations} may be a simulation artifact, arising from ignoring quantum suppression of high frequency modes.

Quantum effects are more important at low temperature, where thermal energy is smaller. 
Furthermore, dissipative forces arising from quantum effects~\cite{volokitin16} may be important at low temperatures~\cite{jentschura16}.

\section{Bonded Rotary Joint Structure and Stability} \sectlabel{stability}

The housing is a faceted piece of crystalline diamond of nominal $C_{2v}$ symmetry~\cite{laidler82} at its extreme energy geometries, i.e., with the rotor adopting a staggered or eclipsed geometry with
respect to the pyramids. The housing contains two mirror plane symmetries and one rotation axis along the mirror symmetry plane axis denoted in \fig{rotor_charges}. Nanodiamonds with dimensions in the nanometer range are routinely synthesized by the detonation method \cite{mochalin11}. Detonation nanodiamonds typically have a highly ordered crystalline core covered by a layer of graphene-like patches or other functional groups that stabilize dangling bonds on the surface. This non-crystalline outer layer is largely a result of the detonation nanodiamond formation and purification processes, and while it is sufficient for chemical stability, it is not necessary. DFT calculations have shown that nanometer-sized diamond structures are stable when the surface is passivated with hydrogen atoms \cite{barnard03, tarasov12}. The hydrogen-passivated facets of the housing shown in \fig{rotor_overview} are predicted to be stable and free from surface reconstruction.


\begin{figure}
\centering  
\includegraphics[width=\widefigwidth]{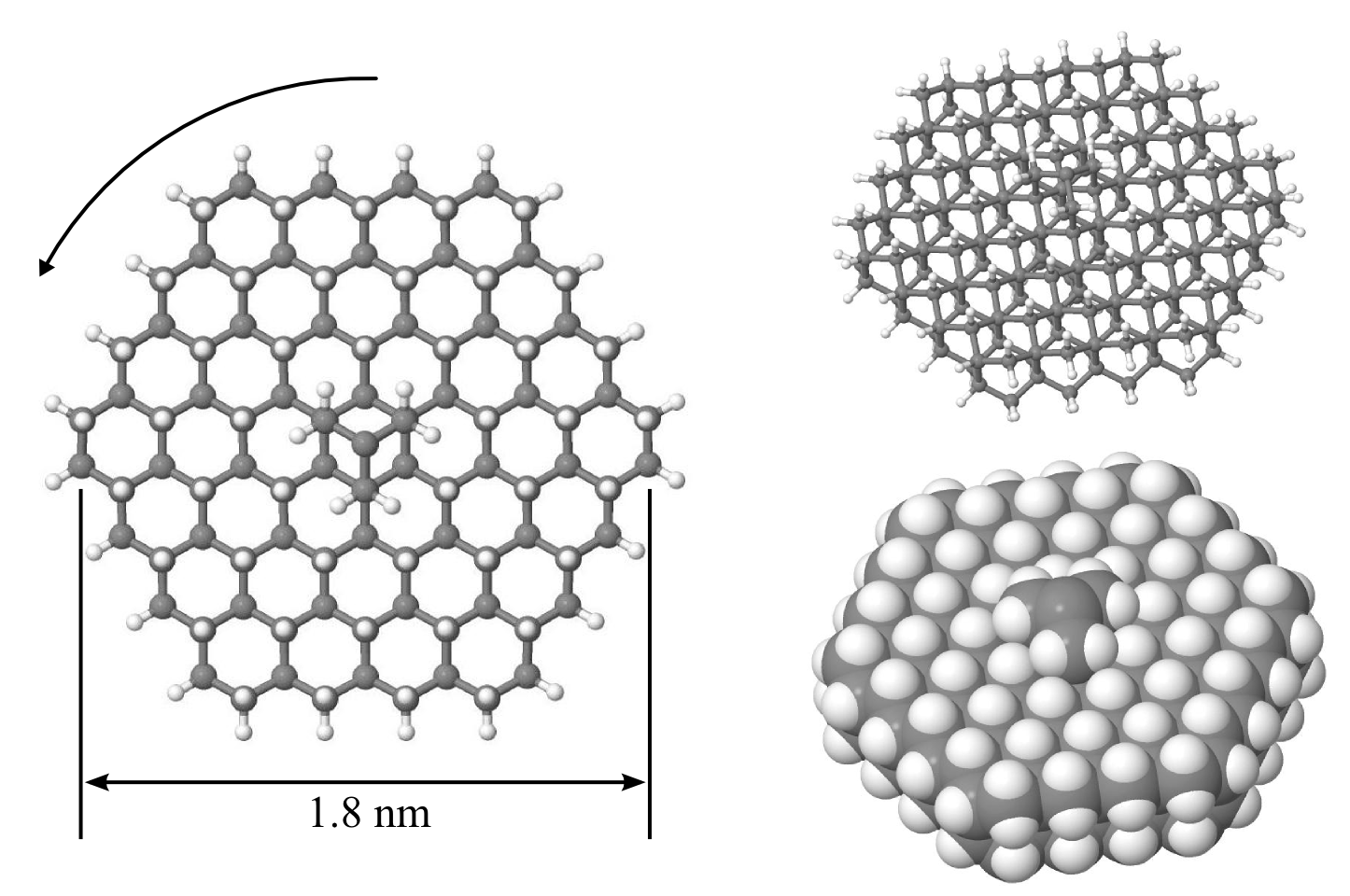}
\caption{The rotor consists of two mirrored and stacked \{111\} layers of hexagonal lattice diamond. The curved arrow indicates the direction of positive angular velocity: counterclockwise when viewed from above, i.e., from along the positive direction of the rotation axis. While the overall shape of the rotor is hexagonal, the atomic structure has threefold rather than sixfold symmetry, as can be seen by close inspection of the atomic structure along the rotor's edges}
\figlabel{rotor_alone}
\end{figure}

The rotor itself is a hexagonally-shaped piece of hexagonal-lattice diamond, or lonsdaleite, as shown in \fig{rotor_alone}. DFT calculations indicate that lonsdaleite would be stable~\cite{nemeth14}, and it has been fabricated~\cite{shiell16}.
As with the housing, the hydrogen-passivated facets of the rotor are predicted to be stable and free from surface reconstruction. 

A ball-and-stick atomic model of the rotor and its two rotary joints is shown in \fig{rotor_charges}. Each rotary joint is composed of four co-linear carbon atoms, similar to the organic molecule 2-butyne. The single bonds at the outside ends of the co-linear chain (e.g., bond C9442-C9443 and bond C9444-C9446) provide for the rotation. Such four-carbon chains have been shown experimentally to function as stable mechanical rotary joints~\cite{shirai05}.
Of the four carbon atoms, the outside two (e.g., C9422 and C9446) are embedded within the framework of the rotating components -- one within the housing and one in the rotor, while the inner two (e.g., C9443 and C9444) are free to rotate. We refer to the inner two carbon atoms as a ``$C_2$ spacer". 

Many proposed applications for molecular machines based on the rotary joint structure are intended to operate in vacuum (e.g. mechanical computation devices). Nevertheless, it is likely that the structure will be stable at room temperature in atmosphere. Several examples of atomically precise mechanical interactions are known to be stable in air, including the self-retraction of nested nanotubes~\cite{zhang13}, self-retraction of graphene flakes~\cite{zheng2008self}, and rotation of acetylenic axles~\cite{shirai05}. In addition, the hydrogen-passivated carbon \{111\} surface is chemically stable in air~\cite{lurie1977diamond,williams2010size}. These experimental results on similar systems strongly suggest that the rotary joint molecule would be stable under ambient conditions.

\section{Density Functional Theory Methods}\sectlabel{DFT_Methods}

Density functional theory (DFT) calculations were used to determine partial charges on atoms around the rotary joint, and to check the rotational potential determined with GROMACS, as described in \sectA{potential details}.

\subsection{Rotor-Pyramid Assemblies}\sectlabel{assemblies}

DFT is computationally demanding so was applied to only a subset of the full structure: the rotor and surrounding pyramids shown in \fig{local_symmetry}, which we call a ``rotor-pyramid assembly''. 

The rotor and housing structure conforms to $C_{2v}$ symmetry~\cite{laidler82} at certain rotor geometries. In the absence of the housing, the rotor-pyramid assembly has significantly higher symmetry at certain rotor geometries. Specifically, defined against the nearest-neighbor H...H interactions between the rotor and pyramids, two symmetry-constrained geometries bring the 12 total pairs of H...H interactions between rotor and pyramid eclipsed and staggered conformations with $D_{3d}$ or $D_{3h}$ symmetries, respectively. \fig{staggered_eclipsed} shows these conformations.
These symmetries allow calculations at significantly higher (i.e., more accurate) levels of theory than is possible for lower-symmetry structures.

\begin{figure}
\centering 
\includegraphics[width=\widefigwidth]{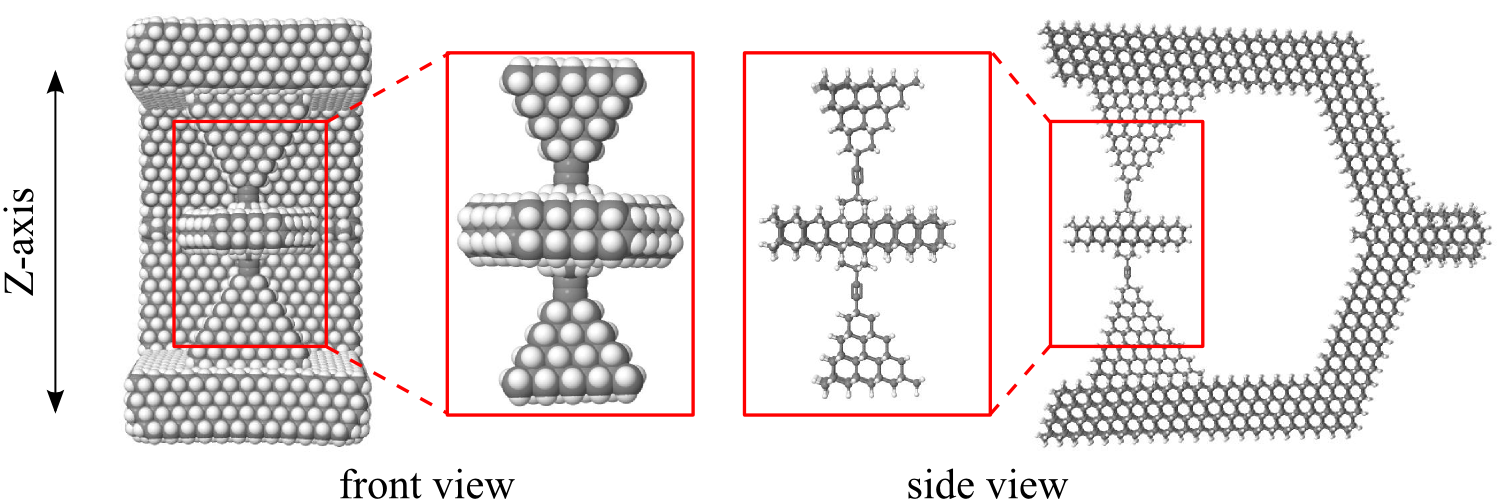}
\caption{Rotor-pyramid assembly used for DFT calculations, shown as insets and in the context of the full structure.}
\figlabel{local_symmetry}
\end{figure}

\subsection{DFT Computations}\sectlabel{DFT details}

Our DFT calculations were performed with Gaussian09, ver.~D.01 \cite{g09} using program-option ``tight" convergence criteria (force criterion $\mbox{RMS} < 1.0 \times 10^{-5}$, density matrix $\mbox{RMS} < 1.0 \times 10^{-8}$), ``ultrafine" grid size (99 radial shells and 590 angular points per shell), and symmetry constraints as applicable to all structures. Optimized geometries for all structures and RESP charges are available on request.

The B3LYP hybrid density functional \cite{becke1993density,stephens1994ab} was employed for all calculations, along with various calculations with the 6-31G(d,p) \cite{hariharan1973influence}, 6-311G(2d,p), and 6-311G(2d,2p) \cite{frisch1984self} Gaussian-type basis sets and D3-type Grimme Dispersion correction~\cite{grimme2010consistent}. 

RESP model \cite{bayly1993well,cornell1993application} charge calculations for the rotor-pyramid assemblies were produced at the B3LYP/6-31G(d,p) level of theory, employing the Merz-Singh-Koller scheme \cite{singh1984approach,besler1990atomic} as a basis for Antechamber script input (program option ``\code{Pop=MK IOp(6/33=2,6/41=10,6/42=17)}").

\subsection{Partial Charges}\sectlabel{partial charges}

\begin{figure}
\centering
\includegraphics[width=\widefigwidth]{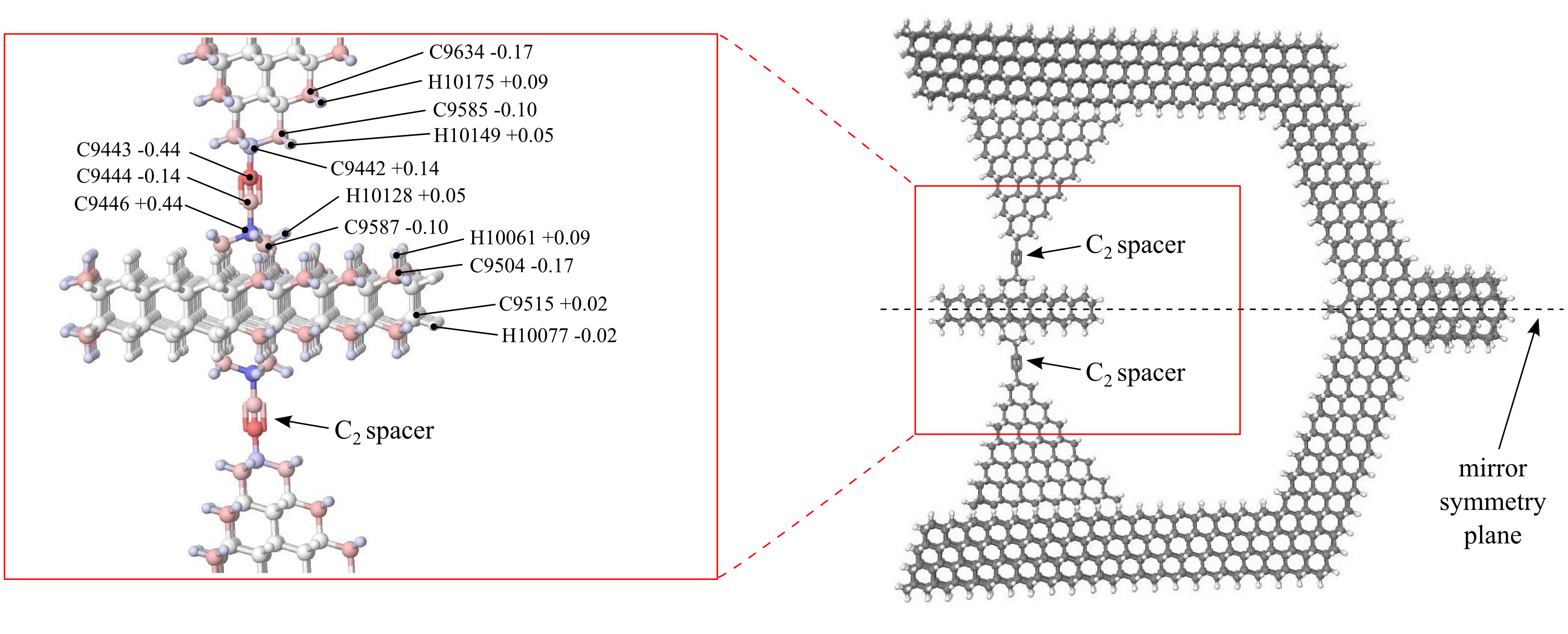}
\caption{Rotor, rotary joints and symmetry plane. Colors indicate restrained electrostatic potential (RESP) charges on atoms in and around the molecular rotary joint. Charge ranges from $-0.44$ to $+0.44$ electron charges (red and blue, respectively). Some atoms in the structure are denoted by type (C or H) and index number, along with their partial charges. These index numbers are those used in the GROMACS structure files, described in \sectA{rotor gromacs details}. 
}
\figlabel{rotor_charges}
\end{figure}

We determined partial charges on atoms near the rotary joint by way of the AMBER Antechamber program~\cite{AMBER2016}. 
\fig{rotor_charges} shows the resulting restrained electrostatic potential (RESP) charges. 
Values for these charges are comparable to published values used for similar molecular joints, e.g. \cite{akimov08}.

\subsection{Rotational Potential Barrier of Rotor-Pyramid Systems}\sectlabel{potential DFT}

\begin{figure}
\centering  \includegraphics[width=\widefigwidth]{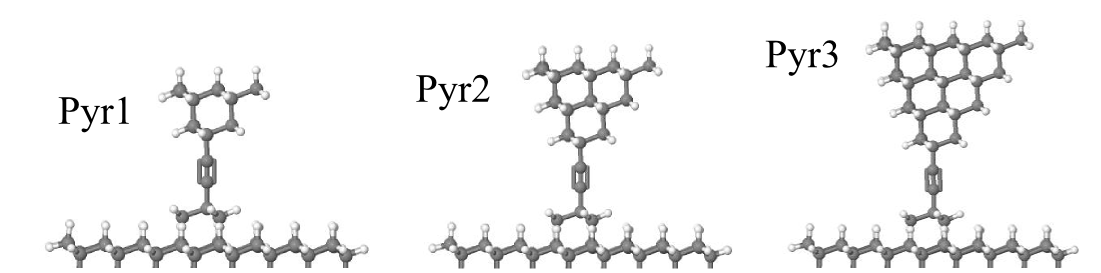}
\caption{Groups of atoms around a rotary joint used for DFT evaluation of rotational potential: Pyr1 (adamantane), Pyr2 (pentamantane), Pyr3 (undecamantane). The bottom layer of atoms is the top of the rotor, and the upper layers are portions of the pyramid.}
\figlabel{rotor_pyr1_pyr2_pyr3_comparison}
\end{figure}

\begin{table}
\begin{center}
\small
\tabcolsep=0.11cm
\begin{tabular}{rcccc}
     		&                & \multicolumn{3}{c}{PyrN-$C_2$-Rot-$C_2$-PyrN} \\
DFT theory level     & $C_{10}$-$C_2$-$C_{10}$  & Pyr1 & Pyr2 & Pyr3   \\
\hline
B3LYP-GD3/6-311G(2d,2p)    & 0.022 & -- & -- & --  \\
B3LYP-GD3/6-31G(d,p)          & 0.030 & 0.063 & -0.006  & 0.096 \\
B3LYP/6-31G(d,p)                  & 0.003 & 0.122 &   0.057 & 0.154 \\
\end{tabular}
\end{center}
\caption{Energy difference, $2V_0$, (in \kJmol) between eclipsed ($D_{3d}$) and staggered ($D_{3h}$) geometries for several molecular joints: $C_{10}$-$C_2$-$C_{10}$, shown in \fig{staggered_eclipsed}, and the groups shown in \fig{rotor_pyr1_pyr2_pyr3_comparison}. The rows show results using different levels of theory, from most (top) to least (bottom) accurate, described in \sectA{DFT details}.
}
\tbllabel{rotational_potential DFT}
\end{table}

The key property of the rotational potential used for our drag estimation is that the barrier, $2V_0$, is relatively small compared to $\BoltzmannConstant T$ as we found to be the case with the GROMACS evaluation of the potential given in \sectA{potential details}. As a check on this relationship, we also evaluated the barrier with the more accurate DFT method. 

To reduce the computational cost of the DFT calculations, we restricted the evaluation to rotor-pyramid assemblies shown in \fig{rotor_pyr1_pyr2_pyr3_comparison}. These assemblies have few enough atoms to allow feasible DFT calculations, while including most of the interactions between the rotor and the housing. We further reduced the cost by only evaluating the highly symmetric eclipsed and staggered conformations of the assemblies described in \sectA{assemblies}. These conformations correspond to the extrema of the potential evaluated with GROMACS in \sectA{potential details}. Evaluating just these two cases is sufficient to compute the potential barrier, i.e., as the difference in energy between the eclipsed and staggered geometries of the joint.

\tbl{rotational_potential DFT} shows the values for the barrier, i.e., $2V_0$, for assemblies with various numbers of atoms in the pyramid. 
The calculated barrier between eclipsed and staggered geometries is well below $\BoltzmannConstant T = 2.5\,\kJmol$ at the temperature we consider.
The rotational barrier does not change noticeably when including more atoms from the pyramid in the calculation, indicating that the barrier is a local phenomenon primarily due to interactions among atoms close to the rotary joint.
The small size of these computed energy barriers ($0.05\,\kJmol$) are in excellent agreement with experimental measurements and theoretical calculations for the rotational barrier around the acetylene linkage in the symmetric 2-butyne molecule~\cite{toyota10}. Diphenylacetylene, a structurally similar molecule with closer H...H interactions than diadamantylacetylene and a preference for maintaining planarity, has a measured rotational barrier of $2.5\,\kJmol$ (calculated to be $3.3\,\kJmol$ at the B3LYP/6-311+G(d,p) level of theory~\cite[Table 2]{toyota10}). At the operational temperature we consider, the barrier for the $C_2$ spacer motif is expected to be inconsequential to the operation of the full assembly.

The barrier computed by GROMACS for the complete structure, $2V_0=0.5\,\kJmol$ from \tbl{rotor parameters}, is larger than that determined by DFT for isolated rotor-pyramid structures ($-0.006$ to $+0.154$ kJ/mol, see \tbl{rotational_potential DFT}). 
Molecular mechanics force fields using RESP charges are of inherently lower accuracy than DFT methods~\cite{wang2000well}. It is therefore not surprising that GROMACS overestimates the rotational barrier since the barrier predicted by DFT is extremely small.
The larger barrier calculated by GROMACS may be due to interactions in the full structure modeled by GROMACS that are not present in the smaller structures modeled by DFT, such as strain induced by the housing or long range interactions between RESP charges. Regardless of the origin of the discrepancy, a barrier height of $0.5\,\kJmol$ is comparable to values reported for smaller molecules with similar rotational structure, e.g., $12\,\kJmol$ for internal rotation in ethane and $0.07\,\kJmol$ for 2-butyne~\cite{pitzer51,toyota10}.

\clearpage


\end{document}